\documentclass{aa}  

\usepackage{graphicx}
\usepackage{txfonts}
\usepackage{lipsum}
\usepackage{subcaption}         
\usepackage{lscape}             
\usepackage{placeins}           
\usepackage{xcolor}
                                
\begin{document}

   \title{
Stellar heavy-element slope index}


%
%
%

   \author{Gerd R{\"o}pke 
           \inst{1}\corrauth{gerd.roepke@uni-rostock.de}       
             \and Friedrich K. R{\"o}pke 
           \inst{2,3}\email{friedrich.roepke@h-its.org}
             \and {David Blaschke}
             \inst{4,5,6}\email{abc@xyz.edu}
        }

   \institute{Institute of Physics, University of Rostock, Albert-Einstein Str. 23-24, D-18059 Rostock, Germany
     \and Zentrum f{\"u}r Astronomie der Universit{\"a}t Heidelberg, Institut f{\"u}r Theoretische Astrophysik, Philosophenweg 12, 
     69120 Heidelberg, Germany
   \and Heidelberger Institut f{\"u}r Theoretische Studien, Schloss-Wolfsbrunnenweg 35, 69118 Heidelberg, Germany
     \and Institute of Theoretical Physics, University of Wroc\l{}aw, Plac Maxa Borna, 50-204 Wroc\l{}aw, Poland
     \and Center for Advanced Systems Understanding (CASUS), Untermarkt 20, D-02826 G{\"o}rlitz, Germany
     \and Helmholtz-Zentrum Dresden-Rossendorf (HZDR), Bautzener Landstrasse 400, D-01328 Dresden, Germany
   }

   \date{\today}

 
  \abstract
  {The distribution of heavy elements in stars is described using a phenomenological approach, in which Lagrange parameters related to temperature and the chemical potentials of protons and neutrons are introduced within a freeze-out concept.
Slope parameters are considered which describe the gross behavior of the distribution of the heavy elements.
Universality and deviations from universality are discussed, and various examples are provided.
These slope parameters may be of interest for characterizing the conditions under which heavy elements form, but the astrophysical sites where heavy elements are produced remain to be determined.}

   \keywords{heavy-element distribution --
                deviations from universality --
                differential-abundance analysis
               }

   \maketitle
\nolinenumbers

\section{Introduction}

\label{sec:Intro}

 Our understanding of the chemical composition of various
 astrophysical objects has advanced significantly in recent
 years. Consequently, chemical evolution -- particularly the cosmic
 origin of elements across the periodic table -- remains an active area
 of research. This study focuses on heavy elements (atomic number $Z >
 30$) beyond the iron peak, where atomic nuclei are most strongly
 bound. Although stellar nucleosynthesis continuously alters the
 distribution of light elements, these burning processes cannot
 produce heavier nuclei. Instead, synthesizing heavy elements requires
 highly specialized conditions, such as hot, dense, and neutron-rich
 environments that enable neutron-capture reactions.

Isotopic distributions are exceptionally well-constrained for both the
Sun \citep{Asplund09,Lodders03} and cosmic rays, while optical
spectroscopy has enabled the determination of chemical abundances for
vast stellar populations. In particular, targeted surveys like the
SAGA project \citep{2024ApJ...976..118G} have mapped unique systems,
including ultra-faint dwarf galaxies, globular clusters, and stellar
streams. Furthermore, tracking specific elements like europium
has proved vital for tracing the initial emergence and chemical
evolution of heavy matter in the early Universe, see, e.g.,
\cite{Wehmeyer:2015sra}.

Crucially, observations from these diverse stellar surveys reveal a
striking phenomenon: the remarkable universality of heavy-element
abundance patterns.  Across vast stellar populations, distinct stars
exhibit nearly identical elemental distribution ratios. This
uniformity persists even within ancient, extremely metal-poor stars
\citep{2010ApJ...724..975R, 2021RvMP...93a5002C,
  2025A&A...704A.282R}. Reviews of this $r$-process universality
\citep{2008ARA&A..46..241S} highlight that even low-metallicity
galaxies in the early Universe display surprisingly uniform
nucleosynthesis signatures, despite the dominant role played by
first-generation stars \citep{2026EPJA...62...45D}.  Strictly speaking,
this chemical universality is defined as a constant ratio between
heavy elements, expressed as $\log \epsilon_Z - \log
\epsilon_Z^\odot = \mathrm{const.}$ or $[Z/Z'] = 1$  across all affected atomic numbers
$Z$ (for the relation to abundances see Sect. \ref{sec:solar} below).
Universality holds for a wide majority of stars including old,
metal-deficient ones. Ultimately, this structural consistency implies
that the astrophysical environments responsible for forming these
heavy elements operate under tightly constrained, universal physical
conditions.

However, notable deviations from this universality exist. A striking
example is found in stars that exhibit a decreasing abundance ratio
with increasing $Z$, resulting in a negative slope for $[Z/Z']$ as a
function of atomic number \citep{2007ApJ...666.1189H,
  2010ApJ...724..975R}. Conversely, other stellar populations display
a positive slope, characterized by an enrichment of heavier nuclei
relative to solar abundances. These $r$-process-enhanced metal-poor
stars can be categorized into $r$-I and $r$-II subclasses
\citep{2005ARA&A..43..531B}, where $r$-II stars exhibit a mild
enhancement in light $r$-process elements alongside a far more
pronounced enhancement in heavier $r$-process elements compared to
their $r$-I counterparts. To disentangle these intricate
nucleosynthesis signatures, recent high-precision
differential-abundance analyses have been deployed
\citep{2025ApJ...994...78S}, also in the investigation of
actinide-boost stars \citep{2025ApJ...984L..43L}.

A fundamental challenge lies in reconciling this observed universality
and its deviations with our current understanding of the astrophysical
sites and physical conditions under which heavy-element
nucleosynthesis proceeds. For instance, models of neutron star mergers
inherently yield highly variable physical conditions, which should
theoretically produce distinct and diverse heavy-element
distributions. Although some works argue that uniformity can
still be maintained under specific conditions
\citep{2025ApJ...990...37K}, there is no consensus on the
physical origin of the observed deviations from universality
\citep[e.g.,][]{2007ApJ...666.1189H}. Conventionally, these anomalies
are attributed to the mixing of material from disparate astrophysical
sites. Along these lines, \citet{2025ApJ...994...78S} invoked multiple
nucleosynthetic sites to explain $r$-I stellar abundances, proposing
three discrete, step-like plateaus to describe the differential
abundances across distinct atomic mass ranges: the $\alpha$- and
iron-peak elements, the lighter $r$-process elements (Sr, Y, Zr), and
the main $r$-process species (Ba to Tm).

Recent parametric frameworks have investigated these nucleosynthetic
conditions. For example, \citet{2026arXiv260117246T} modeled the
physical requirements for the operation of the $r$-process and evaluated
potential astrophysical sites. By analyzing the necessary neutron
flux, they found that a single set of environmental conditions cannot
reproduce the solar $r$-process distribution from the first to the
third peak. Instead, their calculations indicate that fitting the
solar pattern requires a superposition of at least three distinct
components, combining discrete entropies and electron fractions
, see also \cite{2025ApJ...990...37K}.

Similarly, \citet{2025arXiv251113372M} applied a parametric approach
to galactic chemical evolution, testing whether a single set of
$r$-process parameters could account for the enrichment trends of
multiple neutron-capture elements in the Milky Way. They concluded
that no single class of $r$-process events simultaneously explains the
evolution of both light and heavy elements.  This result aligns with
the framework of \citet{2000PhR...333...77Q}, which separated varying
astrophysical conditions into two distinct event classes: a heavy
``H-component'' for elements in $r$-process-rich stars, and a light
``L-component'' for lighter neutron-capture species
\citep{2025ApJ...990...37K}.

In this work, we present an alternative approach by analyzing
heavy-element distributions to characterize astrophysical objects
using a single set of environmental conditions. Specifically, we
determine three Lagrange parameters: the freeze-out temperature and
the chemical potentials of neutrons and protons within the
nucleosynthetic environment.
The corresponding distribution at freeze-out is denoted as initial distribution.
 After freeze-out, reaction kinetics -- in particular
decay processes -- describe the evolution of the initial distribution
at HEFO toward the final, observed distribution.

Rather than invoking a complex, multi-site mixture of discrete
components or rigid abundance plateaus, e.g.,
\cite{2025ApJ...994...78S}, we introduce a phenomenological approach
based on a continuous dependence of the differential abundances on
atomic number $Z$.  Similar to the metallicity, we introduce a slope
parameter as a consequence of the Lagrange parameters which
characterize the freeze-out of the heavy elements in expanding hot and
dense matter. 

While the dominant astrophysical sites for heavy element
nucleosynthesis remain highly debated, this work does not seek to
identify specific stellar environments.
Instead, we adopt a phenomenological approach
to infer the fundamental thermodynamic parameters required to
reproduce the observed heavy element distributions. Because these
heavy nuclei are unaffected by subsequent stellar burning processes,
their abundances serve as pristine relics of the early nucleosynthesis
environment. By extracting the physical conditions present during
their formation, these elemental signatures can also be utilized to
identify co-natal stellar populations, such as moving groups or
stellar streams within the Milky Way, see, e.g.,
\cite{2021ApJ...908...79G}.

This paper is organized as follows. Section~\ref{sec:Lagr} briefly
introduces the Lagrange parameters, followed by illustrative examples
in Section~\ref{sec:examples}.  A comparison with solar abundances is
presented in Section~\ref{sec:solar}, while a differential analysis is
detailed in Section~\ref{sec:Delta}.  The enhancement of actinide
elements is explored in Section~\ref{sec:actinides}.  Finally,
Section~\ref{sec:disc} provides a discussion of the results, and
Section~\ref{sec:concl} presents our concluding remarks.

\section{Lagrange parameters from HEFO and the slope parameter}
\label{sec:Lagr}

The Lagrange parameters, which characterise the distribution of heavy
elements at at freeze-out (initial distribution), determine the
coarse-grained pattern of the observed final distribution.  To
characterise this gross pattern of the final distribution, a slope
parameter is introduced.

\subsection{The heavy element freeze-out (HEFO) concept}

We are considering hot, dense nuclear matter that is expanding and
cooling down.  Below the so-called Mott density, clusters can form
where nucleons become bound into nuclei.  The distribution of these
nuclei is characterized by the mass fraction $X_{AZ}(t) = A
n_{AZ}(t)/n_B$, where $n_{AZ}(t)$ represents the particle number
density of the isotope $\{AZ\}$ (including excited states) and $n_B$
is the total baryon number density.  Reactions between these
constituents of nuclear matter determine the subsequent temporal
evolution of the isotope distribution function.

At high temperatures and densities, the fast forward and backward
reactions balance each other, leading to local thermodynamic
equilibrium (LTE).  A non-equilibrium approach can be formulated using
the Zubarev non-equilibrium statistical operator method, see
\cite{2025Univ...11..323R}.  As long as the nuclear
formation reaction rates remain high, the actual distribution
$n_{A,Z}(t)$ can be approximated by the relevant distribution
\begin{eqnarray}
    \label{nAZ}
    n^{\rm LTE}_{AZ}(t)&\!\!\!\!=&\!\!\!\! R_{AZ}(\lambda_T,\lambda_n,\lambda_p)\,
    \left( \frac{2 \pi \hbar^2}{Am
      \lambda_T}\right)^{-3/2}\nonumber\\ &&\!\!\!\!\times
    \exp\left\{-\frac{E^0_{AZ}(\lambda_T,\lambda_n,\lambda_p)-(A-Z)\,
      \lambda_n(t)-Z \lambda_p(t)}{\lambda_T(t)}\right\}
\end{eqnarray}
where the Lagrange parameters $\lambda_T(t),\lambda_n(t),\lambda_p(t)$
are the non-equilibrium generalizations of temperature $T$ and the
neutron/proton chemical potentials $\mu_n,\mu_p$.  These parameters
are uniquely determined by the known averages of the energy and the
neutron/proton number densities at time $t$.  The Gibbs distribution
(\ref{nAZ}) naturally arises from maximizing the entropy subject to
these given averages.  Here, $E^0_{AZ}(\lambda_T,\lambda_n,\lambda_p)$
is the ground-state energy of the nucleus $\{A,Z\}$.  In a hot and
dense medium, these energies are modified by medium effects such as
mean-field energy shifts and Pauli blocking.  This is in contrast to
the frequently employed nuclear statistical equilibrium (NSE);
treating the system as a mixture of non-interacting nuclei, that only
occasionally undergo reactive collisions, is physically unjustified at
high densities.  The intrinsic partition sum $R_{AZ}$ includes the sum
over all excited states, including contributions from the continuum.
We adopt the specific expression derived in
\citet{2025FrASS..1233496B}; further technical details can be found in
\citet{Natowitz:2022npi}.

As hot, dense matter expands, the distribution of heavy elements
freezes out as soon as the surrounding neutron-rich, dense environment
dissipates and LTE is no longer maintained.  At this heavy-element
freeze-out (HEFO), the state of the system is governed by the Lagrange
parameters $\lambda_T$, $\lambda_n$, and $\lambda_p$.  Following HEFO,
the subsequent temporal evolution of the heavy-element distribution is
governed primarily by decay processes.  When analyzing the mass number
distribution $X_A$, $\gamma$ and $\beta$ decays have no effect since
they conserve the mass number $A$.  Conversely, nucleon emission,
$\alpha$ decay, and fission modify the mass number distribution $X_A$,
driving the transition from the initial to the final abundance
pattern.

For light elements up to Fe, fusion processes remain efficient even
after HEFO.  Consequently, local thermodynamic equilibrium is
maintained down to much lower temperatures and densities, allowing the
distribution to be described by the traditional NSE framework down to
temperatures of approximately 0.5--1~MeV \citep{2025ApJ...990...37K}.
Below these temperatures, nuclear reaction network (NRN) codes such as
\texttt{SkyNet} \citep{Lippuner:2017tyn} and \texttt{WinNet} \citep{reichert2023a} are
required to track the chemical evolution of the system.  This
post-processing of the hydrodynamical profiles of supernova explosions
and neutron star mergers represents the standard paradigm for
calculating chemical abundances.

As noted in previous work, the Lagrange parameters at HEFO can deviate
significantly from the freeze-out conditions assumed for NSE in
standard approaches.  Phenomenological values typically yield
$\lambda_T$ in the range of 4--5~MeV, a baryon number density of
approximately 0.01~fm$^{-3}$, and a proton fraction of $Y_p \approx
0.1$
\citep{1987PhLB..185..281R,2025Univ...11..323R,2025FrASS..1233496B}.
Rather than providing a full microscopic model to compute these
nuclear matter states during expansion, we simply highlight that these
parameter ranges regularly emerge in numerical simulations of
core-collapse supernovae, in the crusts of proto-neutron stars, etc.

The HEFO concept stands in contrast to the standard paradigm for the
origin of heavy elements, which assumes an initial distribution of
elements in which almost no heavy elements are present.  In that
conventional picture, heavy elements are built up sequentially via
neutron-capture reactions \citep{Burbidge:1957vc}. Distinct mechanisms,
such as the the $r$ process and the $s$ process, generate different
abundance patterns with characteristic peaks, see
\citet{2021RvMP...93a5002C,2025ApJ...995....2R}.  By contrast, the
HEFO concept assumes that as the hot, dense nuclear matter expands,
certain features of the heavy-element distribution freeze out early at
elevated temperatures and densities.  This early freeze-out
establishes the global, coarse-grained distribution, while finer
structural details develop at later stages.  Although the spatial
mixing of locally distinct trajectories remains possible, our focus
here is on this early process that establishes the baseline initial
inventory of heavy elements.

\subsection{Solar abundances}
\label{sec:solar}

Complete information regarding isotopic abundances $n_{AZ}/n_B$ is
available for the solar system. 
The mass fraction distribution $X_A = A \sum_Z n_{AZ}/n_B$
describes the nucleon distribution as a function of the nuclear mass
number $A$.  Fine-grained features, such as even-odd staggering,
emerge during the late stages after freeze-out and are properly
described by explicit reaction kinetics.  Because we are interested in
the global behavior of the abundance pattern, a coarse-grained
distribution is more appropriate.  We define the coarse-grained
distribution as $\hat X_{\hat A} = \sum_{m=0}^3 X_{\hat A+m}$, where
$\hat A = 0, 4, 8, \dots$.  We also introduce the $\hat
A$-metallicity, $M_{\hat A} = \sum_{\hat A' \ge \hat A} \hat X_{\hat
  A'}$, which quantifies the fraction of matter bound in clusters with
mass numbers $A' \ge \hat A$.  By definition, the normalization
condition is given by $M_0 = 1$.  Using the adopted solar system
abundances from \citet{2021SSRv..217...44L}, the solar coarse-grained
distribution function $\hat X^\odot_{\hat A}$ and the corresponding
metallicity $M^\odot_{\hat A}$ can be computed, see Fig. 1 of  \cite{2025Univ...11..323R}.

Within the HEFO framework, an initial mass distribution $\hat X_{\hat
  A}^{\rm ini}$ is established at the time of heavy-element
freeze-out, uniquely characterized by the Lagrange parameters
$\lambda_T, \lambda_n$, and $\lambda_p$.  Following freeze-out, the
highly excited heavy nuclei undergo various decay processes.
De-excitation and nuclear reactions
alter the underlying isotopic
distribution.  Ultimately, we observe the final distribution of the
heavy elements, denoted by $\hat X_{\hat A}^{\rm fin}$.  Unlike light
elements, which can still participate in ongoing fusion reactions, the
further synthesis of heavy elements becomes negligible once the hot,
dense, and neutron-rich environment dissipates at HEFO.  Our objective
is to reconstruct the initial distribution $\hat X_{\hat A}^{\rm ini}$
from the observed final distribution $\hat X_{\hat A}^{\rm fin}$ and
to infer the corresponding Lagrange parameters $\lambda_T, \lambda_n$,
and $\lambda_p$. We treat these parameters as a phenomenological index
to characterize the physical conditions driving heavy-element
nucleosynthesis.

Because de-excitation ($\gamma$ decay) and $\beta$ decay conserve the
mass number $A$, the mass number distributions $\hat X_{\hat A}$ and
$M_{\hat A}$ remain invariant under these specific reactions.  In
addition to the baseline normalization condition $M_0 = 1$, a key
constraint is provided by the total amount of matter bound in heavy
nuclei, defined here as $M_{76}$ for elements with $A \ge 76$.  Under
fission or $\alpha$ decay, the resulting daughter nuclei typically
remain within this heavy-element regime.  While a minor reduction in
$M_{76}$ occurs due to $\alpha$-particle emission or neutron
evaporation, this effect can be robustly estimated
\citep{2025Univ...11..323R}.  Consequently, $M_{76}$ is a nearly
conserved quantity.  By combining this conservation with the condition
that the slope of the $A$-metallicity (represented by $\hat X_{76}$)
is also nearly invariant under decay processes, the Lagrange
parameters for the solar heavy-element distribution can be uniquely
determined. Following the methodology in \citet{2025FrASS..1233496B},
these are found to be $\lambda_T^\odot = 5.29$~MeV, $\lambda_n^\odot =
940.294$~MeV, and $\lambda_p^\odot = 845.055$~MeV.

\begin{figure*}[ht]
\begin{minipage}[b]{0.45\linewidth}
\centering
\includegraphics[width=\textwidth]{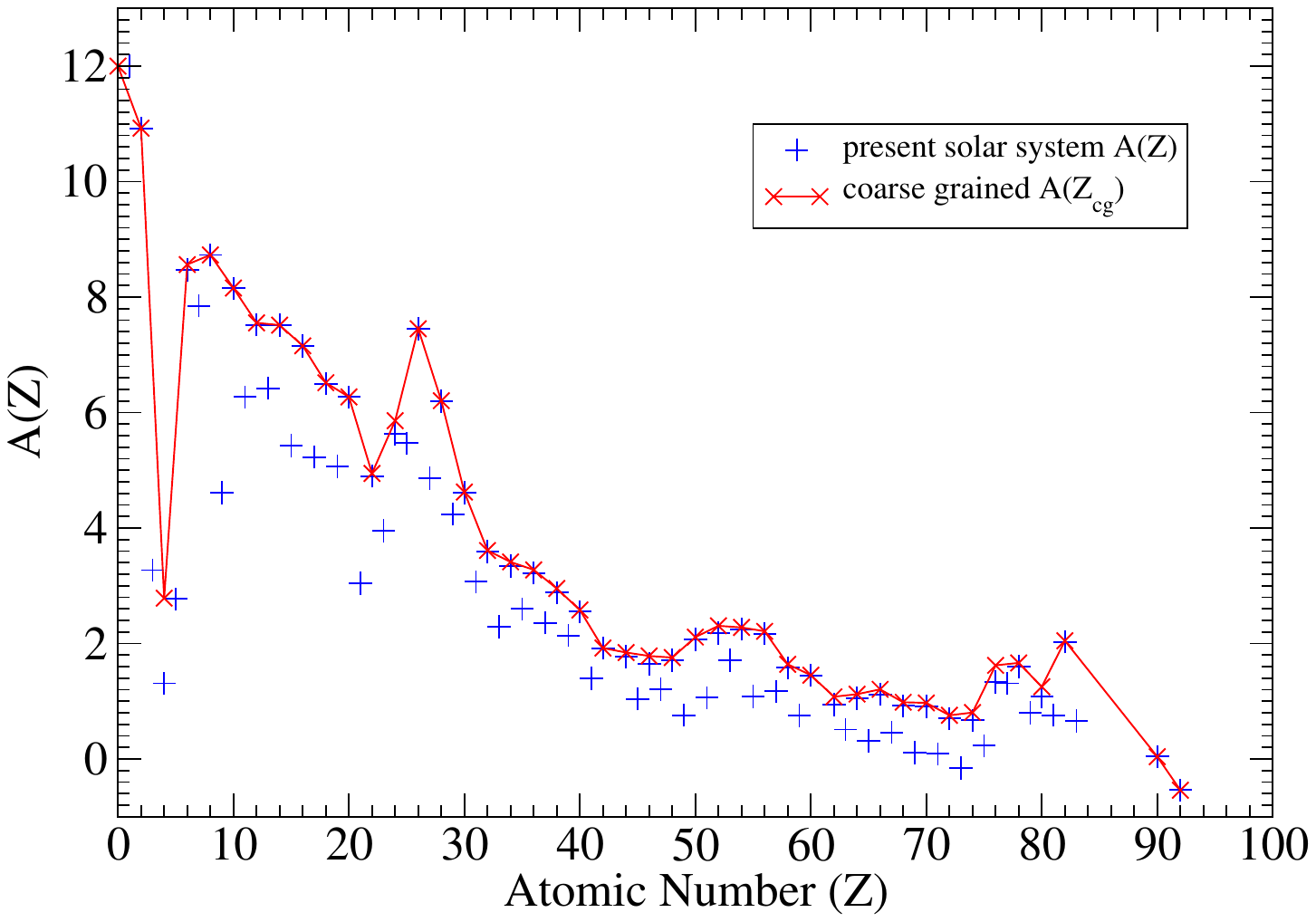}
\caption{Adopted solar abundances of elements $A(Z)$
     \citep{2021SSRv..217...44L} and coarse-grained distribution $\hat
     A(Z_{\rm cg})$.}
\label{fig:YZ}
\end{minipage}
\hspace{0.5cm}
\begin{minipage}[b]{0.45\linewidth}
\centering
\includegraphics[width=\textwidth]{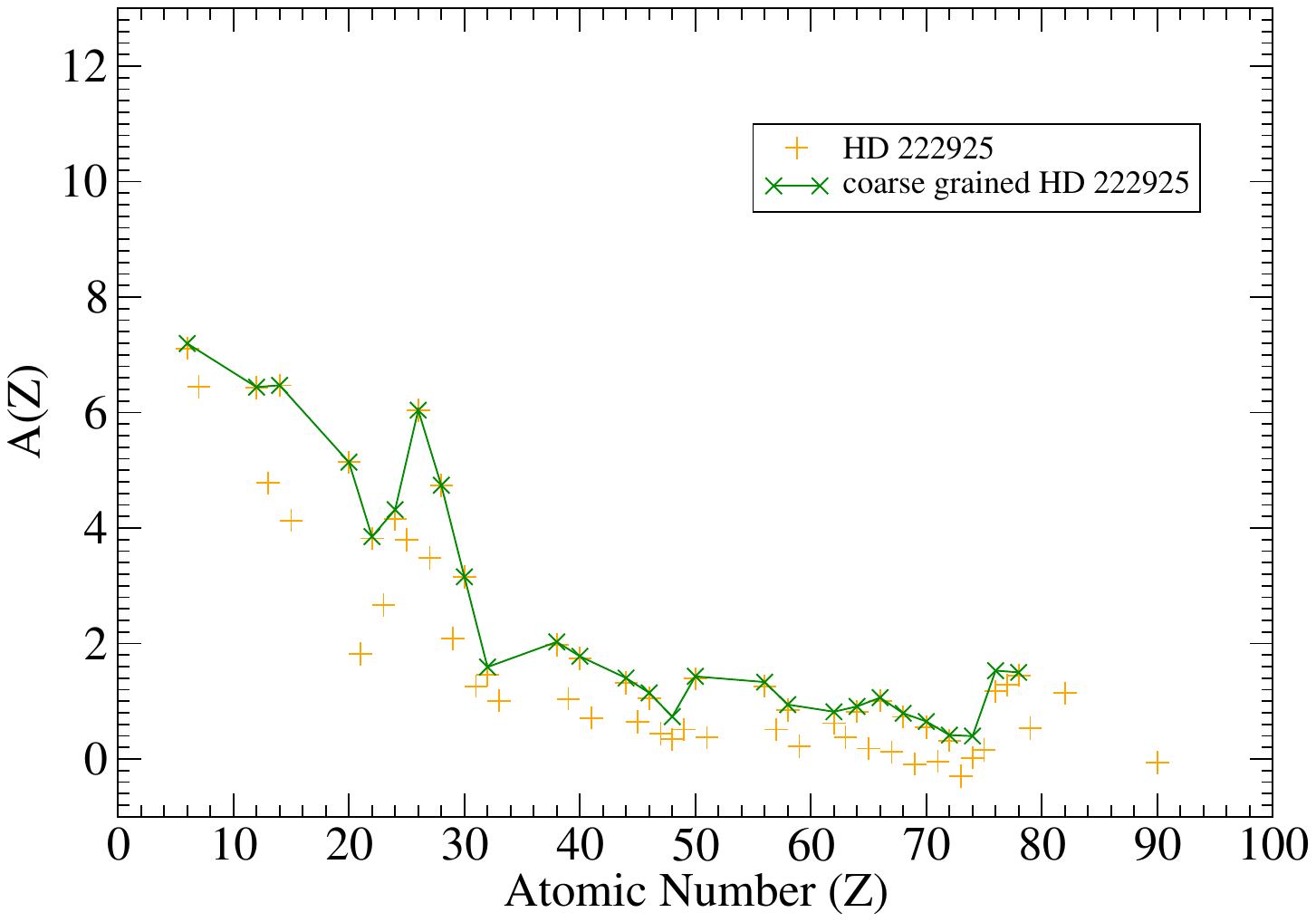}
\caption{Abundances $A(Z)$ \citep{2022ApJS..260...27R} and
     coarse-grained distribution $\hat A(Z_{\rm cg})$ for HD 222925.}
\label{fig:YZstar}
\end{minipage}
\end{figure*}

Instead of the mass number distribution $X_A$, which is highly
convenient for our analytical framework, observational data often
yield elemental abundances defined as $Y_Z = \sum_A n_{AZ}/n_{\rm
  H}$.  A closely related and frequently used observational metric is
the logarithmic abundance scale, $A(Z) = \log \epsilon_Z = \log_{10}
(Y_Z) + 12$.  To map the elemental distribution $Y_Z$ to the mass
number distribution $X_A$, the underlying isotopic distribution must
be known.  For our analysis of the solar system, we adopt the standard
baseline abundances compiled by \citet{2021SSRv..217...44L}, as
illustrated in Fig.~\ref{fig:YZ}.

To smooth out the prominent odd-even staggering effect, we introduce a
coarse-grained elemental distribution defined as $\hat A(Z_{\rm cg}) =
\log_{10}(Y_Z + Y_{Z+1}) + 12$, where $Z_{\rm cg} = 0, 2, 4, \dots,
92$.  A smooth mass fraction distribution $\hat X_{\hat A}$ can then
be mapped directly to $\hat A(Z_{\rm cg})$ by leveraging the robust
empirical relations derived from solar system abundances.  Beyond the
iron peak, the abundance distribution of heavy elements generally
exhibits a decreasing trend.  A prominent exception is the distinct
peak in the lead region, which manifests just before the heavy nuclei
encounter the limits of nuclear stability.  The observed
  abundance-peak in the lead region is usually explained by
  $s$-process reactions.  Part of this peak can already be derived
  from the $\alpha$ decays in the initial distribution in the HEFO
  model \citep{2025Univ...11..323R}, where the actinides Th and U are also produced, along with
  the unstable trans-bismuth elements ($Z >
  83$).

Stellar spectroscopy primarily provides access to these elemental
abundances $Y_Z$.  In observational astrophysics, these abundances are
conventionally expressed relative to solar values using the bracket
notation $[Z/Z'] = \log_{10}(Y_Z / Y_{Z'}) - \log_{10}(Y_Z^\odot /
Y_{Z'}^\odot)$, or more specifically as, for instance, $[Z/{\rm H}]$
for a given element $Z$ relative to hydrogen.  
By mapping each atomic number $Z$ to its corresponding representative
mass number $A$ and scaling these relations directly against the solar
distribution, the exact details of the isotopic distribution become a
secondary effect.  Consequently, extracting the underlying mass number
distribution from raw elemental abundances remains a significant
observational challenge.

\subsection{Abundances of HD 222925 and the slope parameter}

The chemical composition of stars is routinely derived from stellar
spectra.  By analyzing absorption lines, the abundance fraction $Y_Z$,
or equivalently the logarithmic abundance scale $A(Z)$ can be
determined.  As a representative example, we examine the metal-poor
star HD~222925.  HD~222925 is a well-studied, low-metallicity star for
which the elemental abundances $Y_Z$ (or $A(Z)$) have been tracked
across a comprehensive range of atomic numbers $Z$
\citep{2022ApJ...936...84R,2022ApJS..260...27R}.  The observational
abundances $A(Z)$ are presented in Fig.~\ref{fig:YZstar}, alongside
our calculated coarse-grained distribution $\hat A(Z_{\rm cg})$.

\begin{figure}[ht!]
  \centering
  \includegraphics[width=\hsize]{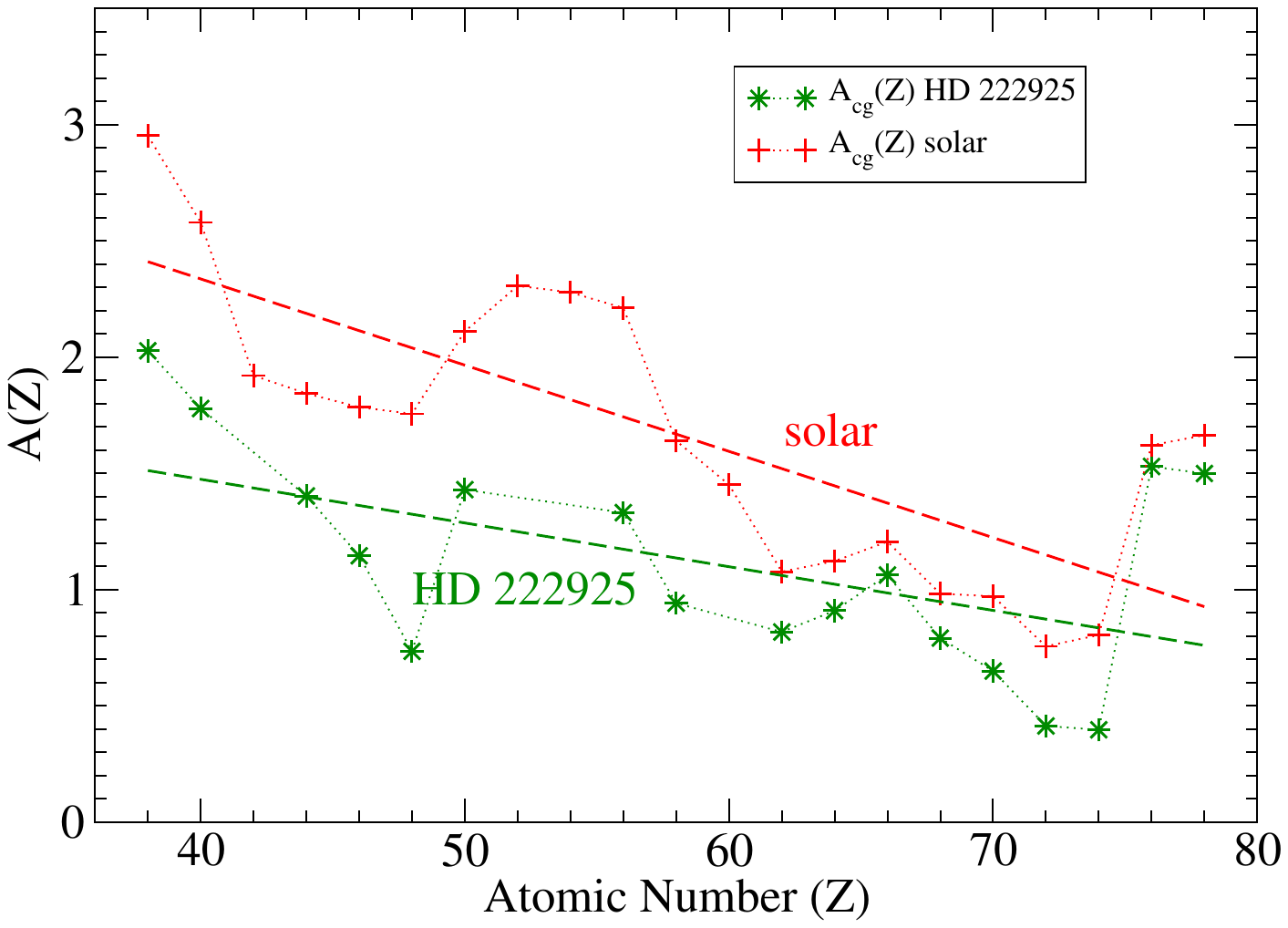}
   \caption{\label{fig:YZstarsolar} Heavy-element coarse-grained
     distribution for HD 222925 in comparison with the solar one.
     Dotted lines serve as a guide to the eye.  The linear
     least-squares fit and the coarse-grained distribution are
     overplotted. Solar: $A(Z)=3.821-0.0371\,Z$; HD 222925:
     $A(Z)=2.227-0.01881 \, Z$}
\end{figure}

When compared with the solar abundance distribution
(Fig.~\ref{fig:YZ}), the coarse-grained distribution for HD~222925
displays a remarkably similar pattern.  However, the entire
distribution is shifted to lower values, reflecting the low
metallicity of the star ($[{\rm Fe/H}] = -1.46$).  Furthermore, the
distribution appears somewhat tilted relative to the solar pattern;
for instance, the values of $\hat A(Z_{\rm cg})$ for $Z_{\rm cg} = 76$
and $78$ are closer to each other than those in the range $40 \le
Z_{\rm cg} \le 50$.  This feature is illustrated more clearly in
Fig.~\ref{fig:YZstarsolar}, which directly compares the heavy-element
coarse-grained distributions of both the Sun and HD~222925.  Our
analysis is restricted to $Z_{\rm cg} \ge 38$, where observational
data are available.

Figure~\ref{fig:YZstarsolar} also displays a linear least-squares fit
to the data points.  The bin corresponding to $Z_{\rm cg} = 82$ is
excluded from the fit, as this region primarily contains $s$-process yields and remnants of
the $\alpha$ decay of trans-lead elements.  The resulting slope for
the solar distribution is $-0.0371$, whereas the slope for HD~222925
is significantly flatter at $-0.0188$.  Naturally, because the
observational uncertainties of individual stellar abundances are
relatively large, these derived slope values have large error bars.

\begin{figure}[ht!]
  \centering
  \includegraphics[width=\hsize]{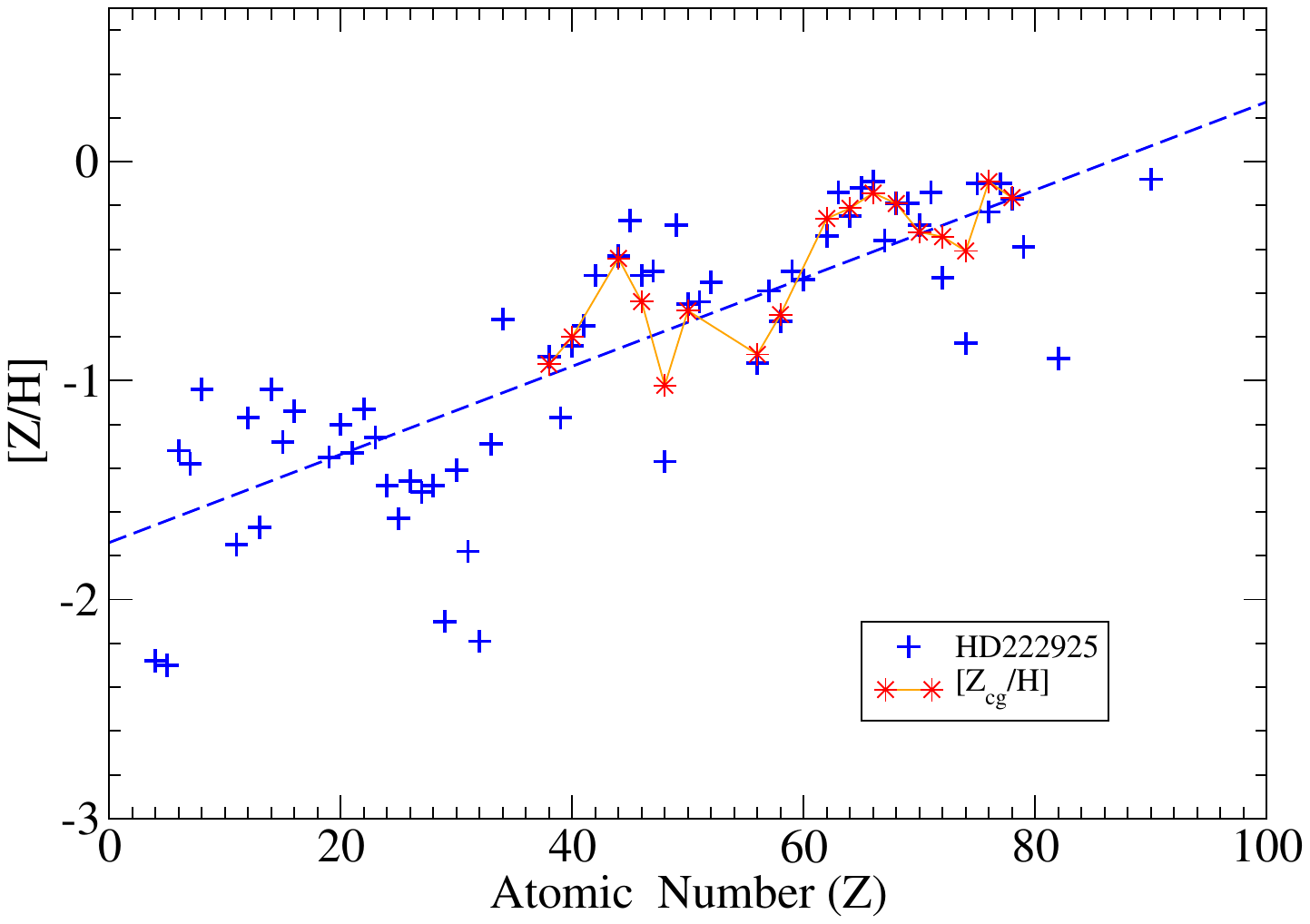}
   \caption{\label{fig:XH222} Ratios $[Z/{\rm H}]$ for HD 222925 given
     in \citet{2022ApJS..260...27R}.  The coarse-grained distribution
     is also shown, dotted lines to guide the eyes.  A least square
     fit is also shown: $[Z/{\rm H}]=-1.74+0.02013 \,Z$.}
\end{figure}

To evaluate these variations more precisely, we examine the abundance
ratio $[Z/{\rm H}]$ relative to the Sun, using data compiled in
Table~3 of \citet{2022ApJS..260...27R}.  
The full
dataset is presented in Fig.~\ref{fig:XH222}.
Analogously, we introduce a coarse-grained differential abundance
ratio, defined for the heavy elements as $[Z_{\rm cg}/{\rm H}] =
\log_{10}(\hat Y_{Z_{\rm cg}} / \hat Y^\odot_{Z_{\rm cg}})$, $\hat
Y_{Z_{\rm cg}}=Y_Z+Y_{Z+1} $, which is also plotted in
Fig.~\ref{fig:XH222} where data permit.  Both abundance ratios reveal
a clear upward trend with increasing atomic number $Z$.  A linear fit
to the complete dataset with $Z \ge 30$ yields $[Z/{\rm H}] = -1.74
+ 0.0201 Z$.  This trend agrees reasonably well with a linear fit
applied exclusively to the available coarse-grained bins, representing
the difference between the two standalone linear fits shown in
Fig.~\ref{fig:YZstarsolar}.
 
\subsection{Universality}
\label{sec:univ}

The concept of universality in the rapid neutron-capture ($r$-process)
elemental abundances was comprehensively reviewed by
\citet{2008ARA&A..46..241S}.  Historically, remarkably identical
heavy-element abundance patterns have been observed across a wide
variety of stars, including extremely metal-poor objects.  Strictly
speaking, an identical abundance pattern implies that the absolute
scaling remains constant, such that $[Z/{\rm H}] = \text{const.}$, or
equivalently, the relative ratios between heavy elements $Z$ and $Z'$
mirror solar ratios, leading to $[Z/Z'] = 0$, see
\cite{2010ApJ...724..975R,2021RvMP...93a5002C}.  This observed
universality of the heavy-element distribution has been interpreted as
a signature of common conditions at which heavy elements are formed.
The overall ubiquity of the rapid neutron-capture mechanism confirms that 
it is a widespread cosmic phenomenon.

However, as growing observational archives provide increasingly
detailed chemical abundance profiles for a diverse stellar population,
clear deviations from strict universality are emerging.  A clear
example is found in HD~222925, whose abundance ratios $[Z/{\rm H}]$
were presented in Fig.~\ref{fig:XH222}.  Evidently, the criteria for
universality ($[Z/{\rm H}] = \text{const.}$) are not satisfied for the
heavy elements.  We explore additional examples of such variations in
Section~\ref{sec:examples}.

Deviations from universality have also been evaluated recently by
\citet{2025FrASS..1233496B}.  Such deviations from a single, uniform
pattern suggest that the heavy elements are synthesized under diverse
physical conditions.  The primary objective of our phenomenological
approach is to characterize and parametrize these variations
systematically.  To achieve this, we map the observed heavy-element
distribution to three characteristic Lagrange parameters, effectively
assigning a unique phenomenological index to each star.

It is a main question of the present work to show whether a
parametrization within the HEFO concept is possible.  This is a
phenomenological approach.  It can help to identify the conditions at
which the formation of the heavy elements occurred.  The question
about the origin of universality and, within a more general view, the
deviation from universality, falls outside the scope of this paper.
Different approaches to explain universality and deviations are given
in Sect. \ref{sec:disc}, where also some astrophysical candidate sites
are discussed, where these conditions can be found.

\subsection{The relative slope}

Our primary objective is to define a quantitative metric to
characterize deviations from strict universality.  While strict
universality means that $[Z/{\rm H}] = \text{const.}$, stellar
observations generally reveal a systemic dependence on the atomic
number $Z$.  To first order, this dependence can be parametrized via a
linear relation for the heavy elements:
\begin{equation}
\label{eq:dz}
    [Z/{\rm H}]= c_Z(Z_0)+d_Z (Z-Z_0).
\end{equation}
This behavior was illustrated empirically for HD~222925 in
Fig.~\ref{fig:XH222}.  The intercept parameter, $c_Z(Z_0)$, represents
the baseline metallicity scale at a given reference element $Z_0$,
analogous to conventional trackers like $[{\rm Fe/H}]$ or, for pure
$r$-process content, $[{\rm Eu/H}]$.  Conversely, the slope parameter,
$d_Z$, constitutes a novel phenomenological index that we evaluate
across various stellar examples in Section~\ref{sec:examples}.  For
the specific case of HD~222925 (Fig.~\ref{fig:XH222}), a linear
regression yields a slope of $d_Z = 0.0201$.  Note that the parameters
$c_Z$ and $d_Z$ refer to a relative distribution (\ref{eq:dz})
determined by the abundances of elements in both the target star and
the Sun.

We map this empirical slope parameter directly onto the physical
framework of the HEFO concept \citep{2025Univ...11..323R}.  Here, the
initial nuclear state is uniquely governed by three Lagrange
parameters: $\lambda_T, \lambda_n$, and $\lambda_p$.  
If we approximate the
coarse-grained initial mass fraction distribution $\hat X_{\hat A}$
within the heavy-element domain ($76 \le \hat A \le 208$) as a linear
profile:
\begin{equation}
    \hat X_{\hat A}=c_A(\hat A_0)+d_A(\hat A-\hat A_0),
\end{equation}
then $c_A(\hat A_0)$ represents a global normalization shift evaluated
at a reference mass number $\hat A_0$.
Concurrently, the slope parameter $d_A$ determines the relative
contribution of the heaviest nuclei in the initial distribution.
Varying the underlying Lagrange parameters naturally shifts the
resulting initial distributions $\hat X_{\hat A}^{\rm ini}$ at
freeze-out.  Consequently, the relative ratio of these mass fractions
with respect to the solar values, $\hat X_{\hat A}^{\rm ini} / \hat
X_{\hat A}^{\odot,\rm ini}$, becomes a sensitive function of the
specific values chosen for $\lambda_i$.  For instance,
\citet{2025FrASS..1233496B} evaluated three distinct coarse-grained
distributions for heavy nuclei ($\hat A \ge 76$) and mapped out the
logarithmic variation $\log(\hat X^{\rm ini}_{\hat A}) - \log(\hat
X^{\odot,{\rm ini}}_{\hat A})$ relative to the solar distribution for
specific choices of $\lambda_i$, see Fig. 2 therein.  By matching both
the shift of $\hat X_{\hat A}^{\rm ini}$ and the slope $d_A$
of the initial state, the corresponding values for $\lambda_i$ can be
mapped out.

\begin{figure}[ht!]
  \centering
  \includegraphics[width=\hsize]{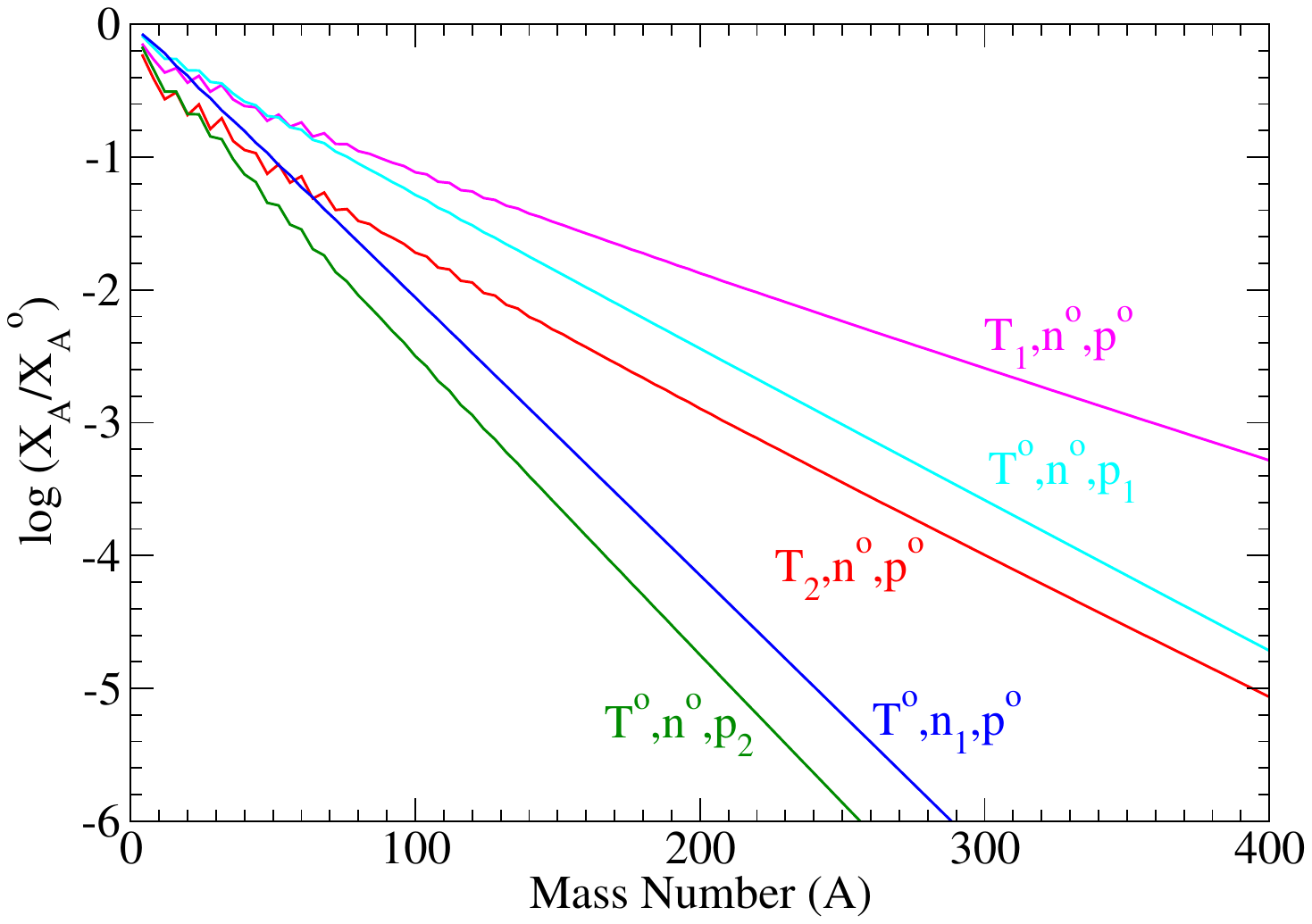}
  \caption{\label{fig:lambda} Relative mass-fraction distribution for
    different Lagrange parameters: $\lambda_T^\odot = 5.2904$~MeV
    ($T^o$), $\lambda^{(1)}_T=5.1$~MeV ($T_1$),
    $\lambda^{(2)}_T=5$~MeV ($T_2$); $\lambda_n^\odot = 940.2941$~MeV
    ($n^o$), $\lambda^{(1)}_n=940$~MeV ($n_1$); $\lambda_p^\odot =
    845.0553$~MeV ($p^o$), $\lambda^{(1)}_p=844$~MeV ($p_1$),
    $\lambda^{(2)}_p=843$~MeV ($p_2$).}
\end{figure}

Figure~\ref{fig:lambda} presents numerical calculations of the initial
ratio $\hat X^{\rm ini}_{\hat A}/\hat X^{\odot,{\rm ini}}_{\hat A}$
under varying configurations of the Lagrange parameters $\lambda_i$,
clearly illustrating how the slope tilts.  The specific parameter sets
evaluated are: $\lambda_T^\odot = 5.2904$~MeV and $\lambda_T=5.1$~MeV,
and $5$~MeV; $\lambda_n^\odot = 940.2941$~MeV and $\lambda_n=940$~MeV;
$\lambda_p^\odot = 845.0553$~MeV, and $\lambda_p=844$~MeV, and
$843$~MeV.  Direct comparison with the solar distribution reveals that
the global normalization shift is primarily governed by the
generalized temperature parameter $\lambda_T$, whereas the slope is
highly sensitive to the proton chemical potential parameter
$\lambda_p$.  As discussed in \cite{2025Univ...11..323R}, the neutron
chemical potential parameter $\lambda_n$ is largely constrained by the
physical Mott density threshold for neutron-rich matter.  In this
work, we focus specifically on extracting and interpreting the
empirical slope parameter, leaving the exact multi-dimensional
inversion to pinpoint definite values for the Lagrange parameters to
future studies.

To link these initial profiles to the final observed mass fraction
distribution $\hat X_{\hat A}^{\rm fin}$, radioactive decay channels
must be accounted for.  For simplicity, we assume that late-stage
neutron evaporation operates under comparable efficiency for both the
solar distribution $\hat X_{\hat A}^{\odot,\rm ini}$ and the target
stars, meaning the initial ratio $\hat X^{\rm ini}_{\hat A}/\hat
X^{\odot,{\rm ini}}_{\hat A}$ is not severely distorted by this
channel.  Conversely, fission and $\alpha$ decay depend strongly on
the original abundance of superheavy elements residing in the tail of
$\hat X_{\hat A}^{\rm ini}$.  Because $\alpha$ decay directly feeds
the final abundance profile in the lead region, we explicitly exclude
this mass range from our slope determination.  The corresponding
impact of fission fragments on the final distribution is detailed
further in Section~\ref{sec:examples}.  To evaluate the slope of the
mass fraction distribution, we adopt the working hypothesis that for
heavy nuclei below the lead region, the differential initial slope
$d_A^{\rm ini}$ derived from $\hat X^{\rm ini}_{\hat A}/\hat
X^{\odot,{\rm ini}}_{\hat A}$ is essentially identical to the final
observed differential slope $d_A^{\rm fin}$ extracted from
$\hat X^{\rm fin}_{\hat A}/\hat X^{\odot, \rm fin}_{\hat A}$.

Finally, we must relate the elemental abundance slope $d_Z$ directly
to the mass fraction slope $d_A$.  In a first-order
approximation we assume that the structural mapping between these two
quantities mirrors the solar configuration.  Approximatively, we can
use the relationship between the solar distributions $X_A^\odot$ and
$Y^\odot_Z$ to derive $X_A^*$ from $Y_Z^*$ for the target star $(*)$ using the same
relationship.  While the absolute determination of the physical
parameters $\lambda_i$ for individual stellar targets falls outside
the scope of this paper, we only demonstrate the general applicability
of this mapping strategy.

In this work, we focus our analysis on the elemental slope parameter
$d_Z$ defined in Eq.~(\ref{eq:dz}). By evaluating abundances of the
target star ($*$) against the solar distribution ($\odot$), we define
the differential slope metric $d_Z=\Delta d_Z^{*,\odot}$ from the linear
regression of the $[Z/{\rm H}]$ profile according Eq. (\ref{eq:dz}).
Extending this approach, the relative variation between two distinct
stars ($*^1$ and $*^2$) can be defined analogously via their
differential slope variance,
\begin{equation}
    \Delta d_Z^{*^1,*^2}=d_Z^{*^1,\odot}-d_Z^{*^2,\odot}.
\end{equation}
This parameter is connected with
$\Delta [Z/Z']^{*^1,*^2}=\log_{10}(Y_Z^{*^1}/Y_{Z'}^{*^1})-\log_{10}(Y^{*^2}_{Z}/Y_{Z'}^{*^2}
)$ for two stars $*^1,*^2$, without reference to solar abundances.

\section{Examples}
\label{sec:examples}

This study focuses on
heavy elements rather than light elements ($Z \le 30$), as the latter
are predominantly determined by stellar burning processes. The
distribution of these heavy elements is expected to provide insights
into the physical conditions and nucleosynthetic sites of early
stellar formation. Specifically, this work addresses three main areas:
(i) a comparison with solar abundances, particularly for stars
exhibiting the ``light neutron-capture excess'' pattern, (ii)
differential-abundance analysis, and (iii) actinide enrichment and
uranium-rich stars.

The primary argument presented here is that HEFO provides an
approximate description of the global matter distribution as a
function of the nuclear mass number $A$, given that fusion processes
and rapid neutron-capture reactions for heavy elements are
subsequently suppressed. Radioactive decay and related reaction
processes modify the local shape of this distribution in $A$-space,
operating analogously to a diffusion process. Conversely, neutron
evaporation acts via single-step processes in $A$-space. Fission and
$\alpha$-decay further alter the distribution by introducing a
significant transfer of matter across $A$-space. However, the precise
contributions of $\alpha$-decay and fission to the final abundances of
heavy elements remain an open problem due to uncertainties in the
underlying reaction rates.

\subsection{Comparison with solar abundances}
\label{sec:solar}

The solar isotopic distribution provides a natural reference system,
typically quantified using the relative abundance ratios
$[Z/Z']$. Many stars exhibit an elemental distribution pattern for
heavy elements that closely resembles the solar distribution. Even in
low-metallicity stars, this characteristic pattern remains evident
when the distribution is renormalized, see, e.g., Fig.~3 in 
\citet{2021RvMP...93a5002C}. Despite this general uniformity,
notable deviations from this universality have been identified.

\begin{figure}[ht!]
  \centering
  \includegraphics[width=\hsize]{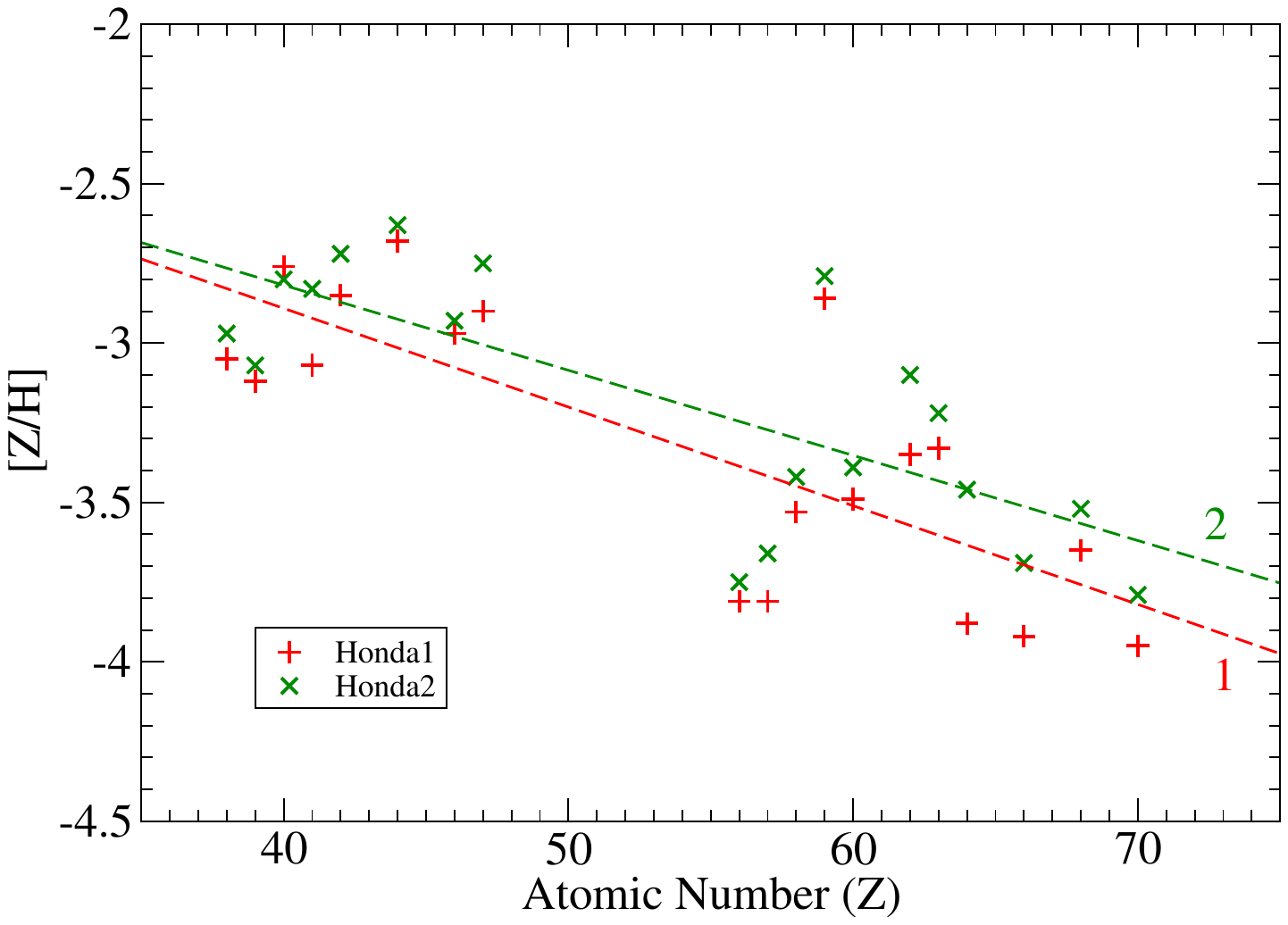}
   \caption{\label{fig:honda} Relative abundances $[Z/{\rm H}]$ of
     HD~88609 (Honda 1) and HD~122563 (Honda 2)
     \citep{2007ApJ...666.1189H}. The linear fits yield $[Z/{\rm
         H}]=-1.653-0.03096 \,Z$ for HD~88609 and $[Z/{\rm
         H}]=-1.75- 0.02671\,Z$ for HD~122563.}
\end{figure}

Significant deviations are observed in the so-called Honda stars
\citep{2006ApJ...643.1180H}. Together with HD~122563, HD~88609 is
often regarded as a prototype of stars exhibiting an unusual ``light
neutron-capture excess'' pattern. HD~122563 is a well-known, extremely
metal-poor red giant in the Milky Way halo. 
It serves as a critical reference object
for studies of early galactic chemical evolution and metal-poor
stellar atmospheres. The observed relative abundances of several heavy
elements for these objects are presented in Fig.~\ref{fig:honda}.

\begin{figure}[ht!]
  \centering
  \includegraphics[width=\hsize]{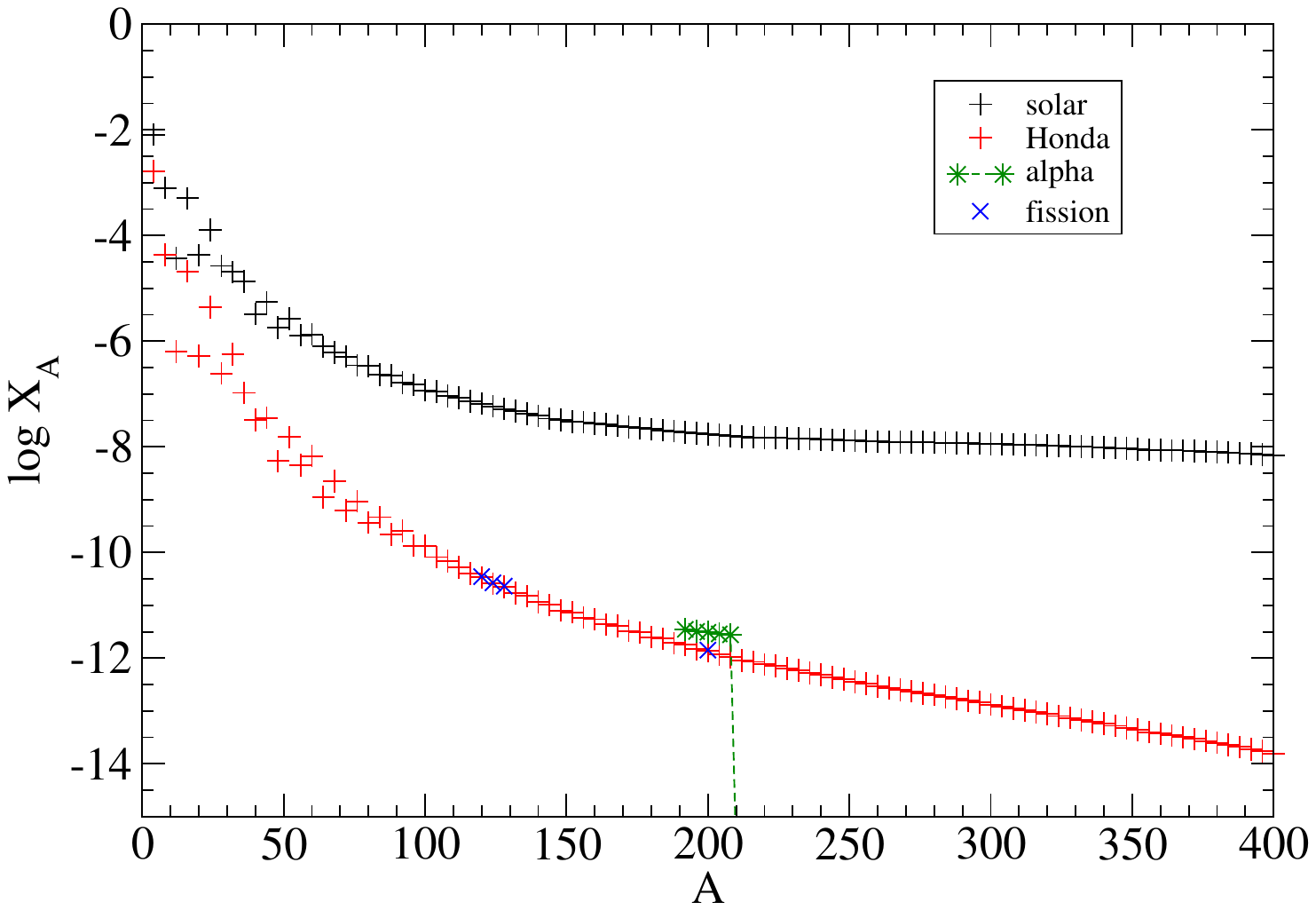}
   \caption{\label{fig:honda3} Initial mass distributions $\hat
     X_{\hat A}^{\rm ini,\odot}$ for the Sun (black) and $\hat X_{\hat
       A}^{\rm ini,Honda}$ for the Honda star (red). The final values
     $\hat X_{\hat A}^{\rm fin,Honda}$ incorporate either exclusive
     $\alpha$-decay for nuclei with $A > 208$ (green, *) or exclusive
     fission for $A > 232$ (blue, x).}
\end{figure}

Figure~\ref{fig:honda} also displays a linear fit derived via the
least-squares method. This fit highlights the deviation from
universality, as evidenced by the negative slope of $d^{\rm Honda}_Z
\approx -0.03$. The HEFO framework accommodates such deviations from
universality through the determination of distinct Lagrange parameters
presented by \citet{2025FrASS..1233496B}. Utilizing the parameters
specified therein ($\lambda^{\rm Honda}_T=4.555$~MeV, $\lambda^{\rm
  Honda}_n=940.944$~MeV, $\lambda^{\rm Honda}_p=842.349$~MeV), the
resulting HEFO mass distribution $\hat X_{\hat A}^{\rm ini, Honda}$ is
shown in Fig.~\ref{fig:honda3} alongside the solar distribution $\hat
X_{\hat A}^{\rm ini, \odot}$.

This initial distribution is expected to evolve post-freeze-out due to
neutron evaporation, $\alpha$-decay, and fission reactions. While
neutron evaporation shifts the distribution toward lower mass numbers
as estimated in \cite{2025Univ...11..323R}, heavier nuclei with
$A \ge 212$ undergo $\alpha$-decay or fission. However, the branching
ratios for these decay pathways remain unknown for most superheavy
isotopes. This uncertainty is less critical for distributions
characterized by a steeply negative slope, as the total mass fraction
contained within superheavy nuclei is small.

The total mass fraction of heavy elements with $A \ge 212$ is
quantified as $M_{212} = \sum_{\hat A \ge 212} \hat X_{\hat A}^{\rm
  ini}$. These unstable nuclei undergo $\alpha$-decay and fission,
ultimately supplementing the final distribution of stable heavy
elements. To evaluate the final distribution $X_{\hat A}^{\rm fin}$,
two scenarios are considered:
\begin{enumerate}
    \item[(i)] All nuclei with $A \ge 212$ undergo successive
      $\alpha$-decay chains to eventually populate the mass range $192
      \le \hat A \le 208$. Accounting for the mass loss from emitted
      $\alpha$ particles, the remaining mass fraction is estimated as
      $\Delta M^\alpha =\sum_{\hat A \ge 212}\hat X_{\hat A}\times
      200/\hat A$. Distributing this mass fraction ($\Delta
      M^\alpha=8.517 \times 10^{-12}$) uniformly over the designated
      $192 \le \hat A \le 208$ interval yields the final configuration
      shown in Fig.~\ref{fig:honda3}, indicating a slight increase in
      final abundances. 
    \item[(ii)] Conversely, assuming that all superheavy elements
      decay via symmetric fission yields a negligible modification to
      the final distribution, which remains imperceptible in
      Fig.~\ref{fig:honda3} (only a small number of $A$ are shown by way of example). 
\end{enumerate}

Stellar environments with a steeply negative distribution slope are
characterized by final heavy-element abundances that are largely
insensitive to nuclear fission and $\alpha$-decay. Such ``baseline
stars'' are of particular interest as reference distributions for
isolating and determining the specific contributions of nuclear
fission \citep{2023Sci...382.1177R}.

Deviations from universality are conventionally analyzed using
multi-component abundance methods; see, for example,
\citet{2018ApJ...856...58H}, which investigates the abundance features
of the metal-poor star HD~94028. For extremely metal-poor stars such
as the halo long-period binary HE~0107-5240 (${\rm [Fe/H]} = -5.56$),
the heavy elements Sr and Ba remain undetected
\citep{2025A&A...704A.238C}.  This star belongs to a binary system
with an estimated period of about 29 years.  Investigating the
distribution of heavy elements in these objects represents an
important avenue for further study.

In contrast, certain stars exhibit a positive slope in their
distribution functions, indicating that heavy components like
actinides are enhanced relative to the solar distribution. The most
comprehensively documented composition of this type is found in
HD~222925 based on data from \citet{2022ApJS..260...27R}, see
Fig.~\ref{fig:XH222}.

Rather than invoking a step-like distribution of varying origins, this
work introduces a continuous slope parameter to characterize the
distribution of heavy elements. Light elements are excluded when
modeling the physical conditions at HEFO. Recent observational
analyses (see, e.g., \citet{2025ApJ...995....2R}) separate the
respective contributions of the $r$- and $s$-processes. The
observational data for key indicator elements such as Sr, Ba, and Eu
(e.g.,  \cite{2005ARA&A..43..531B}) continue to expand
(\citet{2021RvMP...93a5002C,2023ARNPS..73..315L}). 

\subsection{Differential-abundance analysis}
\label{sec:Delta}

Comparing stellar abundances directly to solar values can introduce
systematic uncertainties due to differences in modeling and
observational techniques. A more robust approach utilizes
differential-abundance analysis, where the spectra of multiple stars
are compared line by line.

\begin{figure}[ht!]
  \centering
  \includegraphics[width=\hsize]{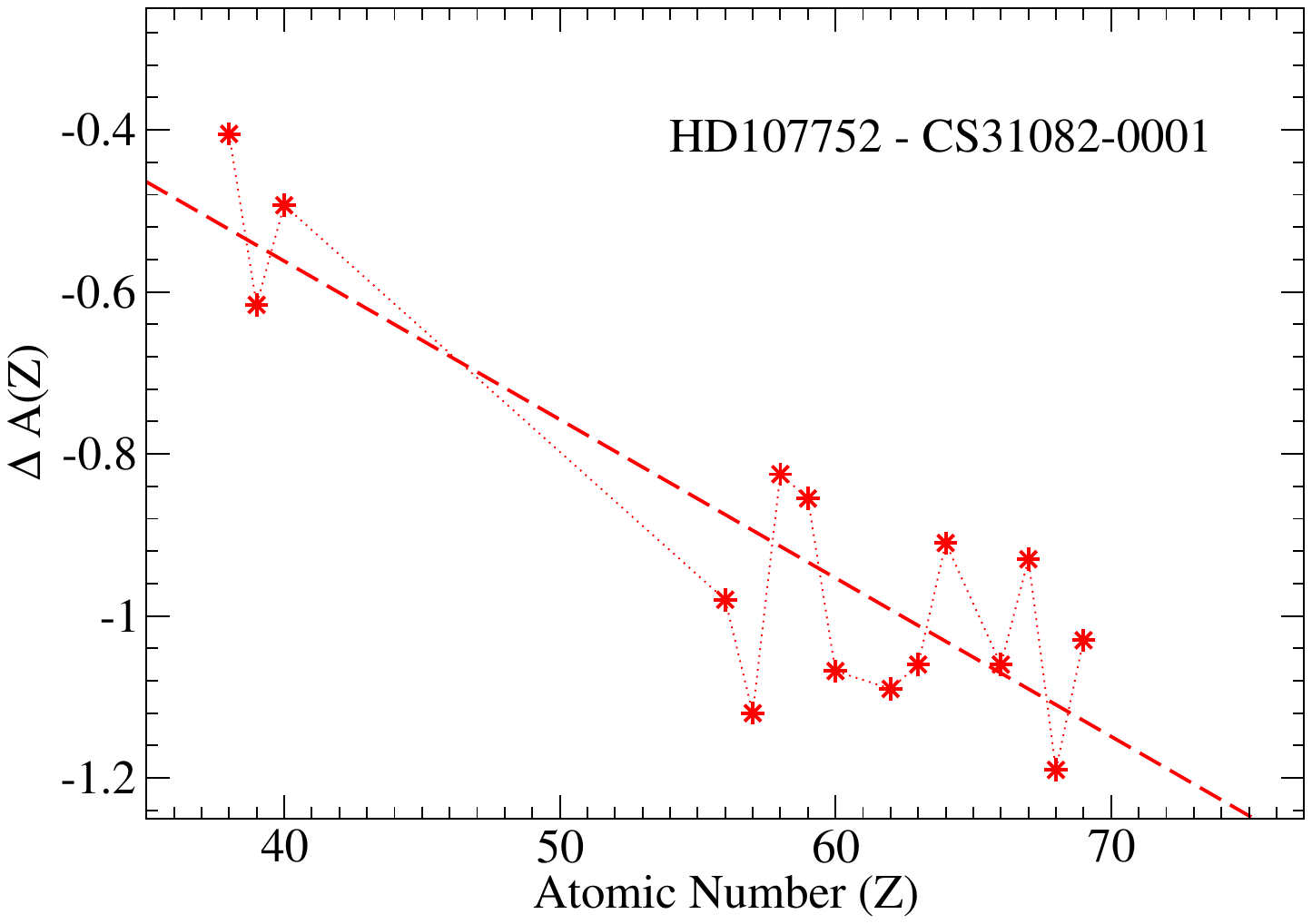}
   \caption{\label{fig:Saraf} Differential-abundance pattern $\Delta
     A(Z)$ of the $r$-I star HD~107752 with respect to the $r$-II star
     CS~31082-0001 \citep{2025ApJ...994...78S}. The linear fit yields
     $\Delta A(Z)=0.2207 - 0.01956\, Z$.}
\end{figure}

\begin{figure*}[ht]
\begin{minipage}[b]{0.45\linewidth}
\centering
\includegraphics[width=\textwidth]{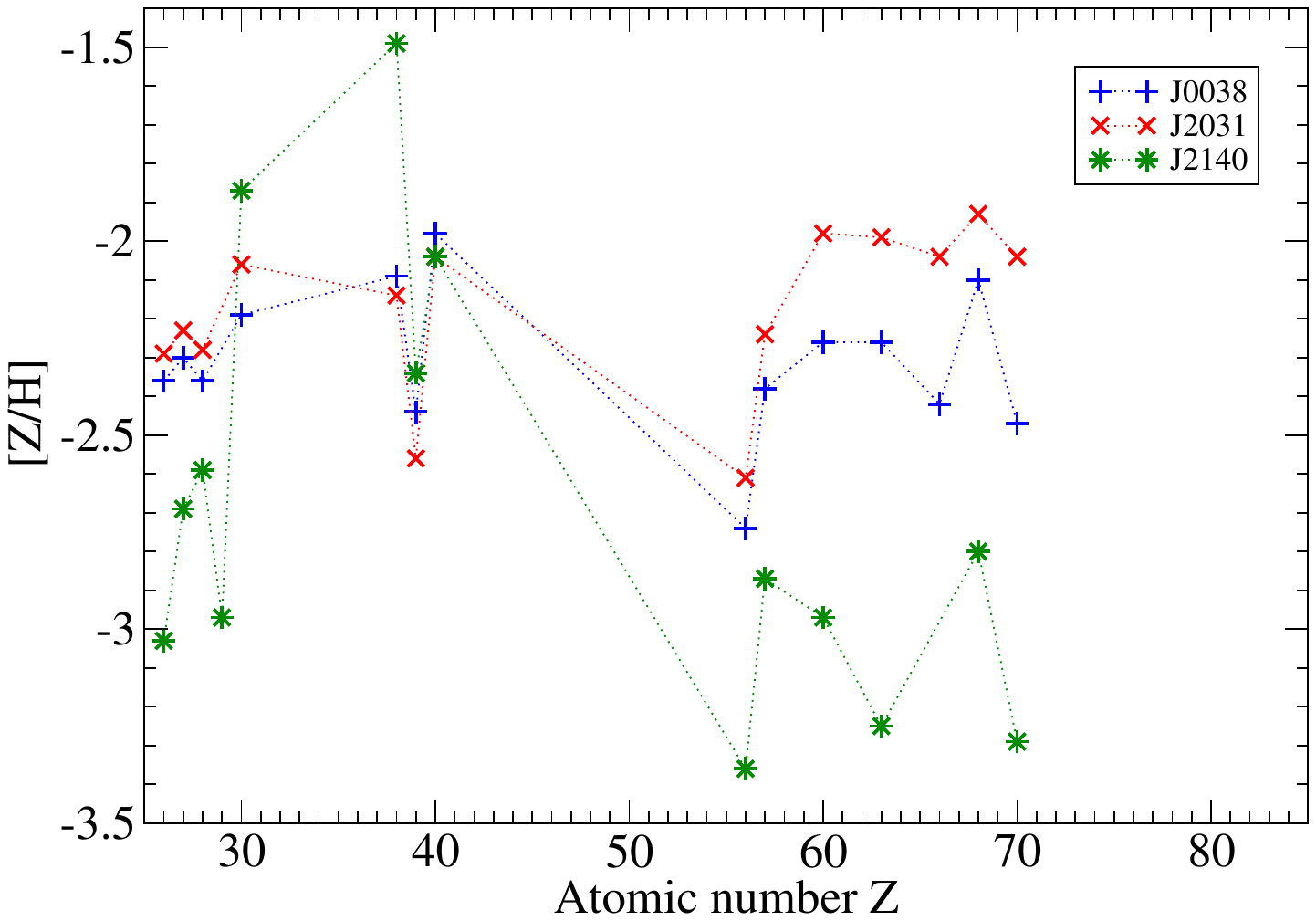}
\caption{Abundances $[Z/{\rm H}]$ for three
     limited-$r$ stars from \citet{2024A&A...688A.123X}.}
\label{fig:Xylakis}
\end{minipage}
\hspace{0.5cm}
\begin{minipage}[b]{0.45\linewidth}
\centering
\includegraphics[width=\textwidth]{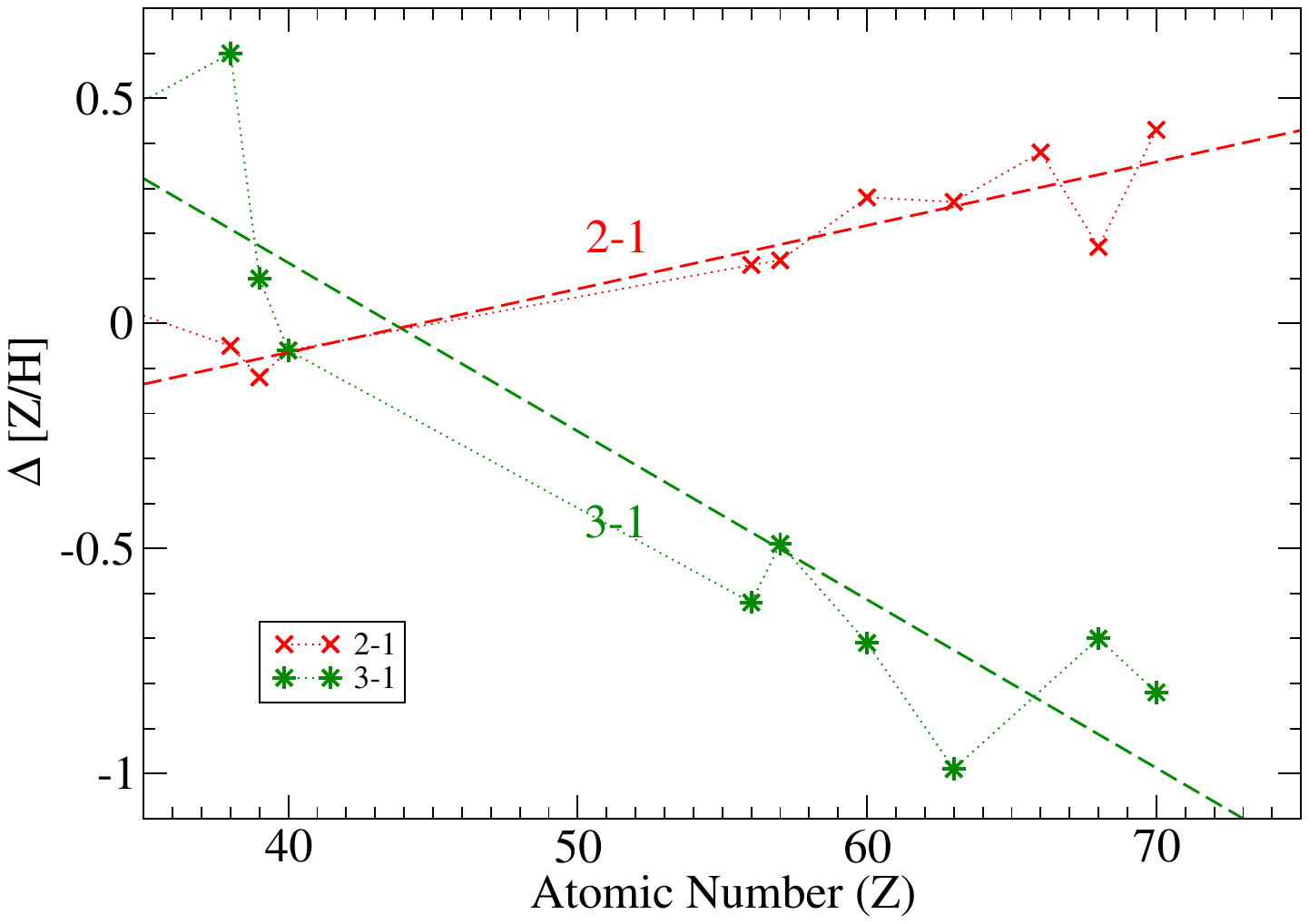}
\caption{Differential abundances $\Delta
     [Z/{\rm H}]$ for three limited-$r$ stars from
     \citet{2024A&A...688A.123X}. The linear fits yield $\Delta
           [Z/{\rm H}]^{*^2, *^1}=-0.6286+0.01411 \, Z$ and $\Delta
           [Z/{\rm H}]^{*^3, *^1}=1.6316-0.03742 \, Z$.}
\label{fig:XylakisDel}
\end{minipage}
\end{figure*}

In a study on the origin of neutron-capture elements,
\citet{2025ApJ...994...78S} performed a differential-abundance
analysis between the $r$-I star HD~107752 and the $r$-II star
CS~31082-0001. The abundance difference as a function of atomic number
is shown in Fig.~4 of
\citet{2025ApJ...994...78S}. Figure~\ref{fig:Saraf} presents a
least-squares linear fit to these data. The abundance difference,
defined as $\Delta A(Z) = A(Z)^{{\rm HD}107752} - A(Z)^{{\rm
    CS}31082-0001} = \Delta [Z/{\rm H}]$, decreases systematically
with increasing $Z$ according to $\Delta A(Z) = 0.2207 - 0.01956\,
Z$. This negative trend reflects the average behavior observed when
comparing populations of $r$-I and $r$-II stars, see Fig.~5 in
\cite{2025ApJ...994...78S}. While \citet{2025ApJ...994...78S}
interpret this pattern using a multi-stage distribution mixture, we
propose that the relationship is well characterized by a continuous,
linear slope.

Another recent example of multi-star comparison is provided by
\citet{2024A&A...688A.123X}, who examined three limited-$r$ stars:
2MASS~J00385967+2725516 ($*^1$, ${\rm [Fe/H]} = -2.48$),
2MASS~J20313531-3127319 ($*^2$, ${\rm [Fe/H]} = -2.38$), and
2MASS~J21402305-1227035 ($*^3$, ${\rm [Fe/H]} = -3.13$). Based on the
reported $[{\rm X/H}]$ values (plotted in Fig.~\ref{fig:Xylakis}), the
resulting differential-abundance analysis is shown in
Fig.~\ref{fig:XylakisDel}. A least-squares fit yields the
corresponding relative slopes.

While \citet{2024A&A...688A.123X} attribute the decline in
neutron-capture abundances at higher atomic numbers to a sequence of
distinct nucleosynthetic events, the astrophysical sites of the
$r$-process remain unconstrained. We suggest that invoking multiple
independent events may not be necessary. Instead, we propose a unified
slope parameter directly determined by the Lagrange parameters that
describe the physical conditions at HEFO.

Differential-abundance analysis isolates the underlying slope and
provides a clearer constraint on the Lagrange parameters
$\lambda_i$. This is evident when comparing Fig.~\ref{fig:Xylakis},
where a linear relationship is difficult to discern, with the
differential trends isolated in Fig.~\ref{fig:XylakisDel}. Accurately
determining this slope is a key prerequisite for calculating the HEFO
Lagrange parameters.

\subsection{Superheavy elements}
\label{sec:actinides}

Within the framework of the HEFO model,
the initial mass distribution $\hat X_{\hat A}^{\rm ini}$ extends
beyond $\hat A=208$. These unstable superheavy nuclei undergo decay,
causing the observed final distribution to terminate in the lead
region. Elements with $Z > 83$ decay via $\alpha$-emission or nuclear
fission; only Th and U persist in the final distribution due to the
exceptionally long half-lives of specific isotopes. While the decay of
various long-lived isotopes is central to cosmochronology,
reconstructing the primordial distribution from current observational
data by accounting for these decay pathways is directly analogous to
our task deriving the initial distribution from the final
heavy-element abundances.

Reconstructing the initial distribution from the observed final
abundances is required because the slope parameter and the
corresponding Lagrange parameters are defined with respect to the
initial distribution. The trans-lead elements present in the initial
distribution (e.g., $^{212}$Po) undergo rapid $\alpha$-decay and
populate the stable lead region of the final
distribution. Concurrently, superheavy elements (e.g., $^{252}$Cf)
undergo fission, yielding daughter products concentrated in the
regions around $Z=44$ and $Z=56$ \citep{nuclei}. As discussed in
Section~\ref{sec:solar}, 
accounting for these processes is critical when deriving the Lagrange
or slope parameters. This calculation remains challenging because the
branching ratios for the various decay channels of superheavy elements
are broadly unconstrained. 
This section solely discusses actinide-boosted,
lead-rich and uranium-rich stars, in which the contributions
of superheavy elements are particularly significant.

\begin{figure}[ht!]
  \centering
  \includegraphics[width=\hsize]{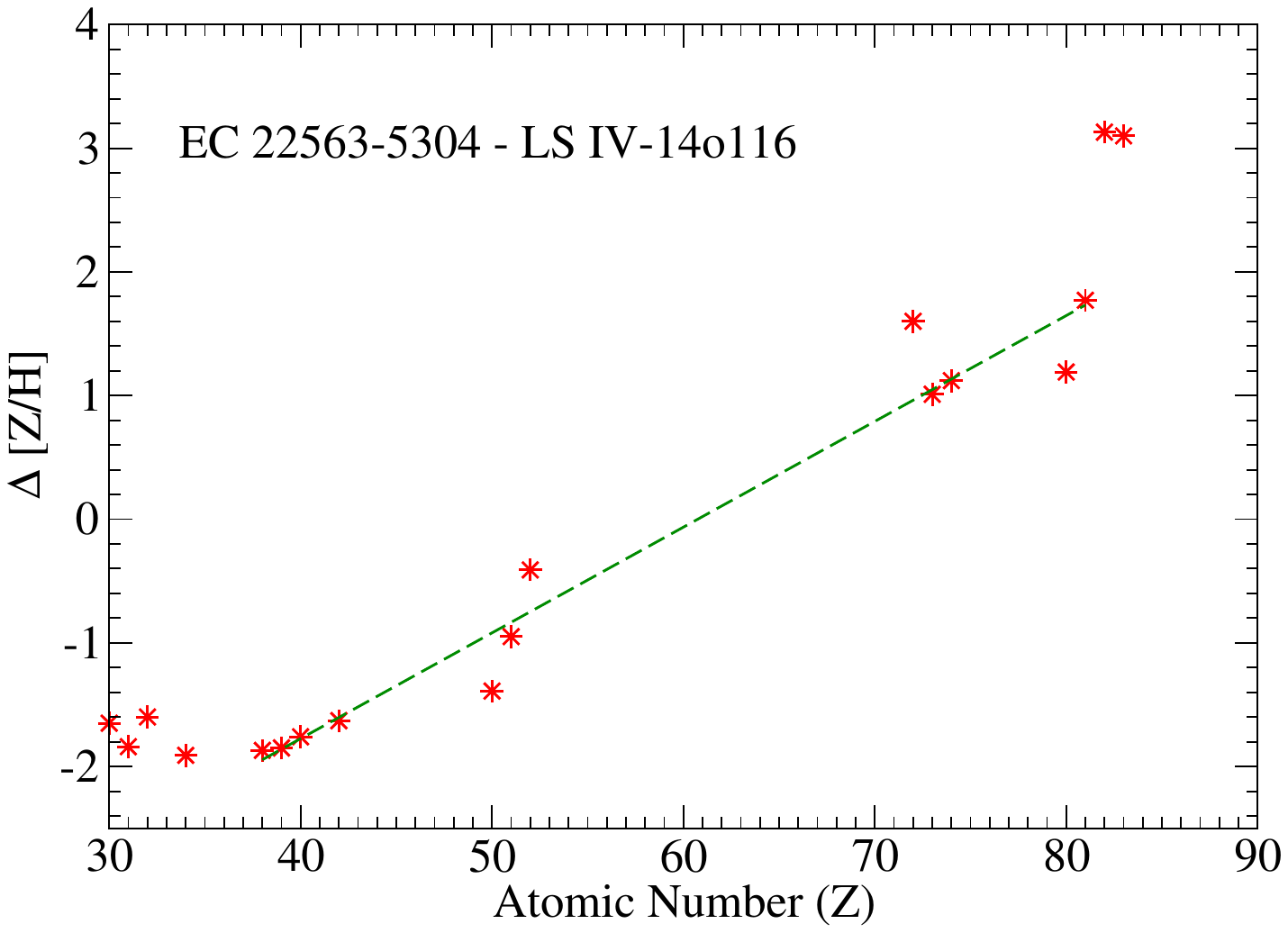}
   \caption{\label{fig:Blei} Abundance difference $\Delta
     \log(n_Z/\sum_i n_i)$ of EC~22536--5304 relative to LS~IV--14o116
     from \cite{2026arXiv260521772D}. Linear fit: $-5.196 +
     0.08552\,Z$}
\end{figure}

A notable example of a star with an elevated lead concentration is
EC~22536--5304; see Table~A3 of \citet{2026arXiv260521772D}. The
abundance difference relative to LS~IV--14o116 displays a clear linear
relationship across the heavy-element range $38 \le Z \le 81$
(Fig.~\ref{fig:Blei}), parametrized as $\Delta \log(n_Z/\sum_i n_i) =
-5.196 + 0.08552\,Z$. Cerium ($Z=58$) is excluded from this fit
because data in the lanthanide region are sparse and carry significant
observational uncertainties. Lead ($Z=82$) and bismuth ($Z=83$) are
also excluded from the least-squares linear fit; their overabundance
relative to the extrapolated linear trend is clearly apparent. 
While the observed abundance in the
  lead region is explained by $s$-process reactions, part of it can
  already be derived from $\alpha$-decays in the initial distribution
  at HEFO, provided that a significant precursor population of
  unstable trans-bismuth elements ($Z > 83$) is present there. 

In general, characterising the heaviest elements (Os, Ir, Au, Pb, Th,
U) observationally is challenging due to their low absolute
abundances. A recent analysis by \citet{2025A&A...704A.282R} concluded
that heavy-element abundance patterns across ten independent
$r$-process sites exhibit remarkably small cosmic dispersion. This
minimal variation implies a high degree of uniformity in $r$-process
yields across diverse astrophysical environments, providing further
evidence for a universal process; see also
Ref. \cite{2008ARA&A..46..241S}. Consequently, the underlying slope
parameter governing the heavy-element distribution appears nearly
identical across these sites.

Nevertheless, significant deviations occur within the heaviest element
regime for certain stellar objects. A prominent example is the
extremely $r$-process-enhanced, metal-poor halo giant CS~31082-001
(${\rm [Fe/H]} = -2.9$), first detailed by \cite{hill2002a}. The
chemical abundances of this uranium-rich object have been extensively
evaluated across multiple studies
\citep{2001ApJ...552L..55Q,2011A&A...534A..60B,2013A&A...550A.122S,2022MNRAS.510.5362E}.
Instead of the step-like behavior shown in the correlation of $[{\rm
    X/Fe}]$ versus $[{\rm Eu/Fe}]$, displayed in Fig.~21 of
\citet{2013A&A...550A.122S}, the abundance distribution can be modeled
via a continuous linear relationship with $Z$, analogous to the
approach in \citet{2025ApJ...994...78S}. Based on the $[{\rm X/Fe}]$
abundances reported by \citet{2022MNRAS.510.5362E}, this trend can be
approximated as $[{\rm X/Fe}] = -0.4 + 0.028\,Z$, which provides an
accurate description of the lanthanide region. A similar linear
increase in the lanthanide region is apparent in the observational
data for the actinide-boost star LAMOST~J122216.85-063345.2, see
Fig.~8 of \citet{2026arXiv260329246J}.

The behavior of the third-peak elements (Hf, Os, Ir, Pt) remains an
area of key interest \citep{2025A&A...693A.294A}, and more precise
  data on their abundances are expected from future UV spectroscopy.
Furthermore, significant lead enhancements have been reported in
objects such as HD~196944 by \citet{2001Natur.412..793V}, see also
\citet{2025ApJ...995....2R}, and broader investigations into stellar
lead abundances have been conducted via the AMBRE project
\citep{2024A&A...690A..97C}.

Observational evidence increasingly supports the role of nuclear
fission in heavy-element nucleosynthesis
\citep{2023Sci...382.1177R}. Specifically, a distinct abundance excess
in the elements Ru, Rh, Pd, and Ag ($Z = 44$--$47$, $A = 99$--$110$)
correlates strongly with the abundances of heavier elements ($63 \le Z
\le 78$, $A > 150$), whereas no such correlation exists for
neighboring elements ($34 \le Z \le 42$ and $48 \le Z \le 62$). This
localized enhancement indicates that fission products of transuranic
nuclei contribute significantly to the observable abundances, implying
that neutron-rich nuclei with mass numbers $A > 260$ are actively
synthesized during $r$-process events
\citep{2023Sci...382.1177R}. Crucially, the inferred fraction of
fission products scales positively with increasing $[{\rm Eu/Fe}]$.

\section{Discussion}
\label{sec:disc}

The origin of heavy elements remains an outstanding and highly debated
problem in nuclear astrophysics. Of particular interest are metal-poor
halo stars, which represent some of the oldest stellar populations in
the Galaxy. These early-generation stars already exhibit significant
enrichment in heavy elements; notably, the observed abundance of
europium at low metallicities cannot be reconciled solely with
conventional neutron star mergers
\citep{Wehmeyer:2015sra}. Consequently, standard nucleosynthetic
pathways struggle to explain the astrophysical sites of these early
$r$-process elements. Alternative proposed environments include
collapsars (rapidly rotating massive stars undergoing core-collapse
supernovae) and magneto-rotational core-collapse events. Rather than
attempting to identify the definitive astronomical site, this work
aims to constrain the physical conditions of heavy-element formation
within a freeze-out framework by determining the corresponding
Lagrange parameters.
A key future objective is to
establish whether these parameter values can be mapped onto specific
physical environments or evolutionary conditions in the early
Universe.

\subsection{Correlation with other stellar properties}

The total heavy-element mass fraction (heavy-element metallicity
$M_{76}$) and the slope of the abundance distribution serve as
phenomenological parameters to characterize a stellar object via
observable properties.  Within the HEFO framework, these empirical
parameters are derived from three underlying Lagrange parameters:
$\lambda_T$, $\lambda_n$, and $\lambda_p$.  These parameters reflect
the physical conditions under which the initial heavy-element
distribution was synthesized, offering insight into the evolutionary
state of the progenitor systems.  An important open question is
whether these nucleosynthetic parameters correlate with other
macroscopic stellar properties.

\begin{figure}[ht!]
  \centering
  \includegraphics[width=\hsize]{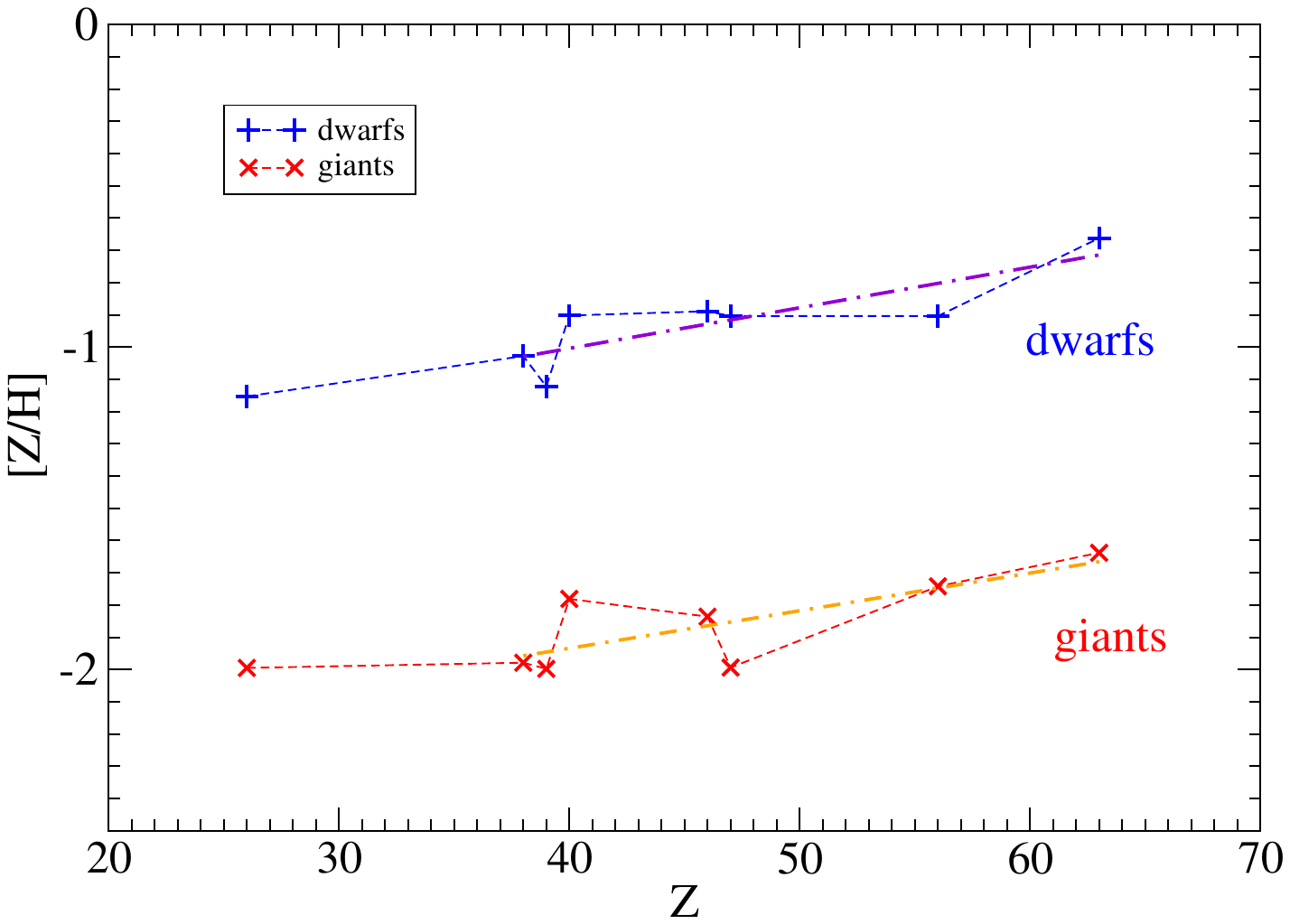}
   \caption{\label{fig:Hansen} Averaged stellar abundances $[Z/{\rm
         H}]$ for halo dwarfs and giants from
     \citet{2012A&A...545A..31H}.  The linear fits are calculated over
     the atomic number range $38 \le Z \le 63$:
     $-1.508+0.0126\,Z$ for dwarfs, and $-2.401+0.01168\,Z$
     for giants.}
\end{figure}

\begin{figure*}[ht]
\begin{minipage}[b]{0.45\linewidth}
\centering
\includegraphics[width=\textwidth]{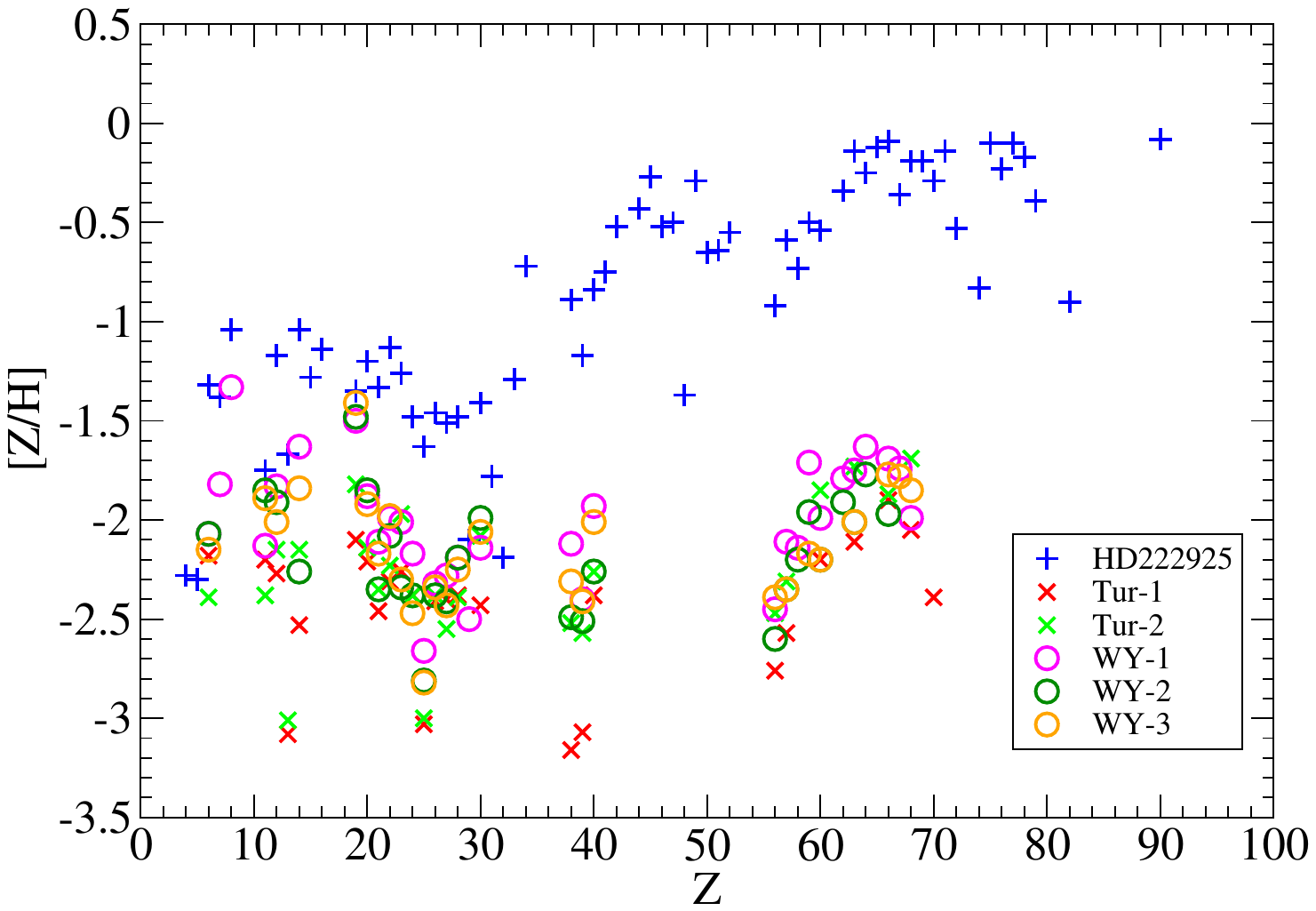}
\caption{Heavy-element abundances $[Z/{\rm H}]$
     for stars associated with the Willka Yaku and Turranburra stellar
     streams, using data from \citet{2026ApJ...998..114W}. The
     abundance pattern of the reference $r$-process-enhanced star
     HD~222925 \citep{2022ApJS..260...27R} is included for
     comparison.}
\label{fig:stream1}
\end{minipage}
\hspace{0.5cm}
\begin{minipage}[b]{0.45\linewidth}
\centering
\includegraphics[width=\textwidth]{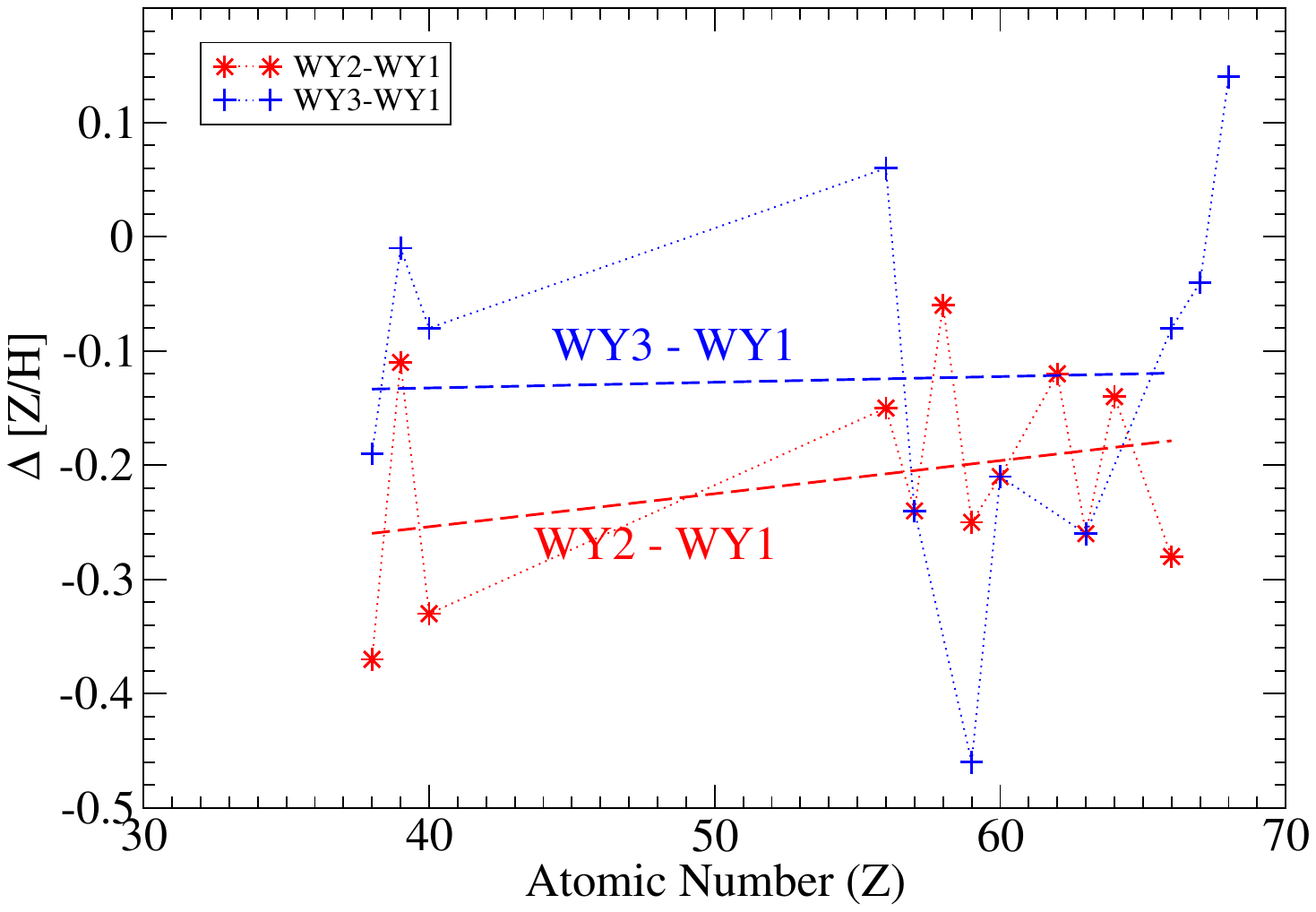}
\caption{Differential-abundance profiles for
     three stars within the Willka Yaku stellar stream based on data
     from \citet{2026ApJ...998..114W}, showing the relative
     differences $\Delta [Z/{\rm H}]^{\rm WY-2, WY-1}$ (linear fit: $
     -0.3697 + 0.002896 \,Z$) and $\Delta [Z/{\rm H}]^{\rm WY-3,
       WY-1}$ (linear fit: $ -0.1525 + 0.000502 \,Z$).}
\label{fig:stream2}
\end{minipage}
\end{figure*}

Such properties include stellar mass, angular momentum, or magnetic
field strength. Furthermore, a significant fraction of these objects
reside in binary systems with close, massive companions. For instance,
the lead-rich object HD~196944 is part of a binary system with an
orbital period of 1294~d, consisting of a companion with an initial
mass of 1.40~$M_\odot$ and a primary star with an initial mass of
0.81~$M_\odot$ \citep{2025ApJ...995....2R}. The Galactic environment
of the stars is also an important variable; low-metallicity stars
(specifically those with low [Fe/H] values) are predominantly found in
the Galactic halo and are traditionally classified as old stellar
populations.

The relationship between the HEFO parameters and stellar mass was
investigated by \citet{2025FrASS..1233496B}, based on the
observational sample from \citet{2012A&A...545A..31H}. That study
evaluated the chemical composition of 42 dwarf and 29 giant field
stars belonging to the Galactic halo as well as the thick and thin
disks. The elemental abundances reported in Tables~C1 and C2 of
\citet{2012A&A...545A..31H} distinguish between dwarf stars (masses
$\approx 0.8~M_\odot$) and giant stars (masses $\approx
1~M_\odot$). Restricting the sample to low-metallicity stars within
the range $-3 \le {\rm [Fe/H]} \le -1$, the averaged stellar
abundances for both groups, originally tabulated in Table~2 of 
\cite{2025FrASS..1233496B}, are presented in
Fig.~\ref{fig:Hansen} alongside their respective least-squares linear
fits.

The linear regressions yield $[Z/{\rm H]} = -1.508 + 0.0126\,Z$
for the dwarf sample and $[Z/{\rm H]} = -2.401 + 0.01168\,Z$ for
the giant sample. While the absolute heavy-element metallicity differs
markedly between the dwarfs and giants, the slopes are nearly
identical ($d_Z^{\rm dwarf} \approx d_Z^{\rm giant} \approx 0.012$). This
indicates that a form of structural universality holds within this
population of low-metallicity stars. However, because this common
slope deviates significantly from zero, these distributions do not
exhibit universality relative to the solar abundance pattern.

The specific Lagrange parameters corresponding to these averaged
distributions were determined by \citet{2025FrASS..1233496B}. The
dwarf sample yields $\lambda^{\rm dwarfs}_T = 4.6391$~MeV,
$\lambda^{\rm dwarfs}_n = 940.875$~MeV, and $\lambda^{\rm dwarfs}_p =
843.873$~MeV, whereas the giant sample yields $\lambda^{\rm giants}_T
= 4.3576$~MeV, $\lambda^{\rm giants}_n = 941.101$~MeV, and
$\lambda^{\rm giants}_p = 843.166$~MeV. The relative abundance
difference between the two populations, expressed as $\Delta [Z/{\rm
    H]}^{\rm dwarfs, giants} = [Z/{\rm H]}^{\rm dwarfs} - [Z/{\rm
    H]}^{\rm giants} \approx 0.9 + 0.001\,Z$, displays a slope that is
consistent with zero, confirming the internal universality shared by
these stellar groups.

\subsection{Stellar streams}

Galactic substructures are frequently mapped using stellar streams,
which represent the tidal debris of disrupted globular clusters or
dwarf galaxies accreted by the Milky Way
\citep[e.g.,][]{2026arXiv260414280S}. 
Chemical tagging 
links these stellar debris fields back to distinct parent systems such
as Sagittarius, $\omega$~Centauri, or the Gaia-Enceladus and Helmi
streams. 
Rather than providing an exhaustive review of the chemical properties
of stellar streams, see, e.g., \cite{2025NewAR.10001713B,2021ApJ...912...52G,2025OJAp....8E..68A,2012MNRAS.425.3188R}, 
we restrict our analysis to a single
representative case study.

Figure~\ref{fig:stream1} displays the $[Z/{\rm H}]$ abundance trends
for stars associated with the Willka Yaku and Turranburra stellar
streams using measurements from \citet{2026ApJ...998..114W}, plotted
alongside the reference $r$-process-enhanced star HD~222925
\citep{2022ApJS..260...27R}. The absolute values of $[Z/{\rm H}]$ do
not exhibit a distinct or uniform trend. Crucially, strict abundance
universality---defined as a constant $[Z/{\rm H}]$ pattern across the
heavy-element regime---is not immediately evident from these data.

As established in Section~\ref{sec:Delta} (cf. Figs.~\ref{fig:Xylakis}
and \ref{fig:XylakisDel}), a direct differential-abundance analysis
removes systematic offsets and facilitates a more robust comparison
between individual stars. Figure~\ref{fig:stream2} applies this method
to the Willka Yaku stream sample from \citet{2026ApJ...998..114W},
plotting the relative abundance difference defined as $\Delta [Z/{\rm
    H}]^{*^2 , *^1} = [Z/{\rm H}]^{*^2} - [Z/{\rm H}]^{*^1}$. Despite
the relatively large observational uncertainties, least-squares linear
fits yield $\Delta [Z/{\rm H}]^{\rm WY-2, WY-1} = -0.3697 +
0.002896 \,Z$ and $\Delta [Z/{\rm H}]^{\rm WY-3, WY-1} = -0.1525 +
0.000502 \,Z$.

The negligible slopes of these differential trends imply that these
three stream members share a common, nearly identical heavy-element
distribution slope.  This internal universality suggests they
originated from the identical environment, with
the slight residual variance falling within acceptable measurement
thresholds.  However, because the absolute abundance differences are
small, the relative error bars remain prominent, underscoring the need
for higher-precision spectroscopic data to confirm these localized
trends.

\subsection{Models and initial conditions}

The temporal evolution of the nuclear distribution function is
governed by reaction kinetic equations. Given the properties of the
constituent species and their respective reaction rates, this
evolution can be computed provided that the initial conditions are
specified. Nuclear reaction networks (NRNs) such as \texttt{SkyNet}
(\cite{Lippuner:2017tyn}) or \texttt{WinNet} (\cite{reichert2023a}) have
been developed to model the chemical evolution of nuclear
systems. While these kinetic models successfully describe the
abundance evolution across diverse astrophysical environments, several
unresolved issues remain:

\begin{enumerate}
    \item[(i)] Nuclear physics constraints remain limited regarding
      the properties of extremely neutron-rich isotopes, including
      their decay rates and reaction cross sections. Ongoing and
      future experimental campaigns are designed to provide tighter
      constraints on these unstable radioactive isotopes.
    \item[(ii)] The applicability of kinetic models based on the
      properties of free nucleons is restricted to low-density
      regimes. Above a critical baryon density of $n_B \approx
      10^{-4}$~fm$^{-3}$, in-medium effects become
      prominent. Specifically, Pauli blocking modifies nuclear binding
      energies, leading to the dissolution of bound states at
      densities below the nuclear saturation density; for details, see
      \cite{2025Univ...11..323R}.
    \item[(iii)] As a system of coupled differential equations, NRNs
      require well-defined initial conditions, for which local
      thermodynamic equilibrium (LTE) is conventionally
      assumed. Hydrodynamic simulations are typically employed to
      model the expansion of hot, dense matter in core-collapse
      supernovae or neutron star mergers. Post-processing with NRNs is
      then performed for temperatures below a critical threshold of
      $T_c \approx 0.5$~MeV, assuming chemical equilibrium holds above
      $T_c$. We question this threshold and propose that
      non-equilibrium distributions emerge within the heavy-element
      regime at significantly higher temperatures. Consequently, we
      utilize relative slope parameters to evaluate these primordial
      initial conditions. The HEFO framework postulates a freeze-out
      temperature on the order of $\sim 5$~MeV, below which a
      non-equilibrium kinetic treatment is strictly required.
\end{enumerate}

Various nucleosynthetic frameworks are used to establish initial
abundance distributions. Several models simulate violent dynamic
phenomena (e.g., core-collapse supernovae, neutron star mergers)
adopting low-metallicity initial compositions. A superposition of
distinct nucleosynthetic processes is frequently invoked to explain
observable distributions, see, e.g., \citet{2025ApJ...990...37K}. A
recent overview of the physical environments governing heavy-element
distributions was provided by \citet{2026arXiv260117246T}, suggesting
that at least three or four distinct physical conditions are necessary
to reproduce observed trends. These frameworks rely on multiple
independent nucleosynthesis sites and subsequent macroscopic
mixing. Similarly, \citet{2014ApJ...797..123H} proposed a combination
of separate nucleosynthetic components to explain stellar abundances
at low metallicity, adopting distinct production mechanisms for
different atomic number intervals.

For instance, the Ceres2025 collaboration demonstrated that current
models for $r$-process sites fail to reproduce the observed third-peak
abundances in these stars \citep{2025A&A...693A.294A}. It remains a
persistent challenge that none of the astrophysical environments
currently under investigation can fully account for the elevated Os,
Ir, and Pt abundances, even when factoring in nuclear physics
uncertainties. Further evaluations of the astrophysical and nuclear
physics conditions at $r$-process sites are detailed by
\citet{2023ApJ...951...30H} for HD~222925. Furthermore,
\citet{2025ApJ...994...78S} concluded that simple dilution of the
$r$-process yield from a single site---such as a neutron star merger,
collapsar, or magneto-rotational supernova---cannot explain the
differential-abundance patterns observed between $r$-I and $r$-II
stars. 
In contrast to these multi-component mixing models, our
framework simplifies this description by characterizing the global
heavy-element distribution within a star using a single set of initial
parameters.

For these initial conditions, we adopt a framework established in the
field of relativistic heavy-ion collisions to model expanding, hot,
and dense nuclear matter. We propose that the freeze-out of heavy
elements occurs during the early stages of the hydrodynamic expansion,
at higher temperatures and densities than those conventionally assumed
in NRN calculations. The coarse-grained, global pattern of the mass
fraction distribution function $X_A(t)$ is established at this
freeze-out stage and is parametrized by Lagrange multipliers. The
fine-grained structure of the observed elemental abundances
subsequently develops via late-stage reactions post-freeze-out, where
the final distribution is determined by neutron evaporation,
$\alpha$-decay, and nuclear fission. While a mixing of different
components representing varied initial conditions remains possible, it
is not uniquely required to explain observed deviations from abundance
universality.

The definitive astrophysical site of heavy-element synthesis remains
unconstrained. Observational evidence confirms that heavy-element
distributions undergo continuous modifications. Radioactive decay of
long-lived actinide isotopes occurs continuously, and dynamic
reactions involving heavy elements have been directly observed in
active stellar phenomena, such as neutron star mergers
\citep{watson2019a}. Physical conditions compatible with HEFO Lagrange
parameters are also known to occur within the crusts of proto-neutron
stars \citep{2025Univ...11..323R}. This raises the possibility that
analogous conditions were attained during the early evolutionary
phases of the Universe.

The early appearance of heavy elements, inferred from the chemical
compositions of extremely metal-poor halo stars, see, e.g.,
\cite{Wehmeyer:2015sra}, presents challenges for the standard paradigm
of homogeneous Big Bang nucleosynthesis (HBBN). In the standard HBBN
framework, which occurs approximately 3 minutes post-Big Bang,
primordial nucleosynthesis ceases after producing hydrogen, helium,
and trace amounts of lithium. This primordial matter subsequently
forms Population~III stars. Consequently, models relying on HBBN must
invoke specialized, energetic early stellar events to generate the
heavy elements observed in the oldest stellar populations.

An alternative pathway is provided by inhomogeneous Big Bang
nucleosynthesis (IBBN). This model posits that massive, high-density,
and high-temperature structures---potentially formed via primordial
phase transitions or within the environments of primordial black holes
prior to standard nucleosynthesis---survived the early expansion
era. These regions can naturally sustain the physical conditions
described by the HEFO freeze-out Lagrange parameters. High-mass
objects in the early Universe have been increasingly identified via
observations with the James Webb Space Telescope
\citep{2023ApJ...947L..24M}, and the survival of primordial black holes
past the 3-minute threshold is discussed by
\citet{Gonin:2025uvc}. Adopting the IBBN framework
offers a mechanism to explain early
heavy-element synthesis, implying that a fraction of the primordial
cosmic matter bypassed the standard chemical constraints predicted by
HBBN models. Detailed discussions of IBBN and its nucleosynthetic
implications can be found in \citep{1987PhLB..185..281R,
  1994ApJ...429..499R,2025Univ...11..323R,
  Gonin:2025uvc,2025FrASS..1233496B,
  2026EPJA...62...45D,2026arXiv260117246T}.

The primary objective of this work is to determine the physical
conditions under which heavy-element distributions form across diverse
stellar environments. To this end, we present a unified
phenomenological framework and evaluate its performance against
several observational case studies. Future spectroscopic observations
are expected to expand the availability and precision of abundance
data for additional chemical elements. While the present analysis is
restricted to a selected sample of well-studied stars with extensive
published heavy-element datasets, a rigorous identification of the
specific astrophysical sites that host these freeze-out conditions
remains the subject of forthcoming investigations.

\section{Conclusions}
\label{sec:concl}

The distribution of heavy elements in stellar atmospheres preserves
critical information regarding the early evolutionary history of
cosmic matter. Because these heavy species are unaffected by stellar
burning processes, their primordial abundance profiles can be linked
to energetic nucleosynthetic phenomena, such as neutron star
mergers. Notably, significant amounts of these heavy elements were
already synthesized in the early Universe. Rather than invoking a
specific microscopic model to explain the onset of abundance
universality and its subsequent deviations, this work provides a
unified phenomenological framework. Assuming a heavy-element freeze-out
(HEFO) scenario, we introduce a continuous slope parameter, $d_Z$, to
characterize the heavy-element abundance distributions observed in
different stars. This parameter provides valuable constraints on the
physical environments and conditions characterizing early stellar
systems during heavy-element synthesis. Furthermore, while
differential-abundance analysis of stellar absorption spectra
represents an ideal methodology for determining this slope parameter,
the late-stage decay pathways of superheavy elements must be
rigorously taken into account. Much like absolute metallicity
measurements, this continuous slope parameter can serve as a robust
diagnostic tool to identify distinct stellar subgroups, such as
kinematically linked stellar streams within the Milky Way.

\begin{acknowledgements}
The authors would like to thank Ian Roederer for his helpful comments.
  G.R. acknowledges a stipend from the Foundation for Polish Science
  within the Alexander von Humboldt programme under grant
  No. DPN/JJL/402-4773/2022. The work of F.K.R.\ is supported by the
  Klaus Tschira Foundation and by the Deutsche Forschungsgemeinschaft
  (DFG, German Research Foundation) -- RO 3676/7-1, project number
  537700965.  F.K.R.\ acknowledges funding by the European Union (ERC,
  ExCEED, project number 101096243). Views and opinions expressed are
  however those of the authors only and do not necessarily reflect
  those of the European Union or the European Research Council
  Executive Agency. Neither the European Union nor the granting
  authority can be held responsible for them.  
\end{acknowledgements}

\end{document}